\documentclass[bibyear]{aa}  
\usepackage[utf8]{inputenc}
\usepackage{graphicx}
\pdfoutput=1

\usepackage{amsmath}
\usepackage{amsfonts}
\usepackage{txfonts}
\usepackage{epstopdf}
\usepackage{times,epsfig}
\usepackage{mathrsfs}
\usepackage{natbib}
\usepackage{amssymb}
\bibpunct{(}{)}{;}{a}{}{,}
\usepackage[english]{babel}
\usepackage{longtable}
\usepackage[dvipsnames]{xcolor}
\usepackage[]{hyperref}
\usepackage{deluxetable}
\usepackage{float}

\colorlet{myurlcolor}{violet}
\colorlet{myallcolor}{MidnightBlue}
\hypersetup{
  linkcolor  = myallcolor,
  citecolor  = myallcolor,
  urlcolor   = myurlcolor,
  colorlinks = true,
}
\newcommand{\Msun}{~M_\odot}

\newcommand{\kms}{\rm ~km~s^{-1}}

\ifdefined\Hy@StartlinkName
\def\addOneNestingLevelStartLink{%
  \gdef\Hy@StartlinkName##1##2{%
    \sbox0{\Hy@StartlinkNameOrig{##1}{##2}}\usebox0
    \global\let\Hy@StartlinkName\Hy@StartlinkNameOrig%
  }%
}
\def\addOneNestingLevelEndLink{%
  \gdef\pdfendlink{%
    \sbox0{\pdfendlinkOrig}\usebox0%
    \global\let\pdfendlink\pdfendlinkOrig%
  }%
}
\let\Hy@StartlinkNameOrig\Hy@StartlinkName
\let\pdfendlinkOrig\pdfendlink
\else
\let\addOneNestingLevelStartLink\relax
\let\addOneNestingLevelEndLink\relax
\fi

\authorrunning{Karamehmetoglu et al.}
\begin{document}

    \title{The luminous and rapidly evolving SN~2018bcc:}
    \subtitle{Clues toward the origin of Type Ibn SNe from the Zwicky Transient Facility}

   \author{E.~Karamehmetoglu \inst{1,9} \href{https://orcid.org/0000-0001-6209-838X}{\includegraphics[scale=0.5]{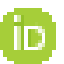}}
        \and C.~Fransson \inst{1}
        \and J.~Sollerman \inst{1} \href{https://orcid.org/0000-0003-1546-6615}{\includegraphics[scale=0.5]{ORCIDiD_icon16x16.eps}}
        \and L.~Tartaglia \inst{1}
        \and F.~Taddia \inst{1}
        \and K.~De \inst{2}
        \and C.~Fremling \inst{2}
        \and A.~Bagdasaryan \inst{2}
        \and C.~Barbarino \inst{1} \href{https://orcid.org/0000-0002-3821-6144}{\includegraphics[scale=0.5]{ORCIDiD_icon16x16.eps}}
        \and E.~C.~Bellm \inst{5}
        \and R.~Dekany \inst{3}
        \and A.~M.~Dugas \inst{2}
        \and M.~Giomi \inst{8}
        \and A.~Goobar \inst{6}
        \and M.~Graham \inst{2}
        \and A.~Ho \inst{2}
        \and R.~R.~Laher \inst{4}
        \and F.~J.~Masci \inst{4}
        \and J.~D.~Neill \inst{2}
        \and D.~Perley \inst{2}
        \and R.~Riddle \inst{3}
        \and B.~Rusholme \inst{4}
        \and M.~T.~Soumagnac \inst{7}
    }
	\institute{The Oskar Klein Centre, Department of Astronomy, Stockholm University, AlbaNova, 10691 Stockholm, Sweden.\\ \email{emir.k@astro.su.se}
	\and Division of Physics, Mathematics, and Astronomy, California Institute of Technology, Pasadena, CA 91125, USA.
	\and Caltech Optical Observatories, California Institute of Technology, Pasadena, CA  91125, USA.
	\and IPAC, California Institute of Technology, 1200 E. California Blvd, Pasadena, CA 91125, USA.
	\and DIRAC Institute, Department of Astronomy, University of Washington, 3910 15th Avenue NE, Seattle, WA 98195, USA.
	\and Oskar Klein Centre, Department of Physics, Stockholm University, SE 106 91 Stockholm,  Sweden.
	\and Benoziyo Center for Astrophysics, Weizmann Institute of Science, Rehovot, Israel.
	\and Humboldt-Universit{\"a}t zu Berlin, Newtonstra\ss{}e 15, 12489 Berlin.
	\and Department of Physics and Astronomy, Aarhus University, Ny Munkegade 120, DK-8000 Aarhus C, Denmark.}

   \date{Received XXX; accepted XXX}

\abstract
{Supernovae (SNe) Type Ibn are rapidly evolving and bright (M$_\text{R,peak}$ $\sim-19$) transients interacting with He-rich circumstellar material (CSM). SN~2018bcc, detected by the ZTF shortly after explosion, provides the best constraints on the shape of the rising light curve (LC) of a fast Type Ibn.}
{We used the high-quality data set of SN~2018bcc to study observational signatures of the class. Additionally, the powering mechanism of SN~2018bcc offers insights into the debated progenitor connection of Type Ibn SNe.}
{We compared well-constrained LC properties obtained from empirical models with the literature. We fit the pseudo-bolometric LC with semi-analytical models powered by radioactive decay and CSM interaction. Finally, we modeled the line profiles and emissivity of the prominent \ion{He}{i} lines, in order to study the formation of P-Cygni profiles and to estimate CSM properties.}
{SN 2018bcc had a rise time to peak of the LC of $5.6^{+0.2}_{-0.1}$~days in the restframe with a rising shape power-law index close to 2, and seems to be a typical rapidly evolving Type Ibn SN. The spectrum lacked signatures of SN-like ejecta and was dominated by over 15 He emission features at 20 days past peak, alongside Ca and Mg, all with V$_{\text{FWHM}} \sim 2000~\text{km}~\text{s}^{-1}$. The luminous and rapidly evolving LC could be powered by CSM interaction but not by the decay of radioactive \element[][56]{Ni}. Modeling of the \ion{He}{i} lines indicated a dense and optically thick CSM that can explain the P-Cygni profiles.}
{Like other rapidly evolving Type Ibn SNe, SN~2018bcc is a luminous transient with a rapid rise to peak powered by shock interaction inside a dense and He-rich CSM. Its spectra do not support the existence of two Type Ibn spectral classes. We also note the remarkable observational match to pulsational pair instability (PPI) SN models.}


   \maketitle

\section{Introduction}

Supernovae (SNe) Type Ibn and IIn are thought to be strongly interacting with the circumstellar material (CSM) surrounding their progenitor stars. The shocks created by the interaction between ejecta and CSM convert kinetic energy into radiation \citep[e.g.,][]{Chevalier2017}. The interaction also modifies the spectral features of SNe Ibn and IIn, as well as the luminosity and duration of their light curves (LCs) (where one characterization of duration is rise time, often defined as the interval between the estimated explosion epoch and the peak of the LC). The -n suffix represents the fact that Ibn and IIn SNe display strong and relatively narrow lines of He and H, respectively; presumably a direct signature of this interaction \citep[e.g.,][]{Smith2016}.

Unlike the longer rise times of the more common Type IIn SNe, several Type Ibn SNe have very fast LCs with rise times of a few days followed by a similarly rapid decline \citep{Hosseinzadeh2017}. Such rapidly evolving LCs make "fast Type Ibn SNe" (rise times $\lesssim 1$ week) a member of the mysterious group of short-duration extragalactic transients \citep{Whitesides2017}. This part of the phase space of optical transients is both interesting and yet poorly constrained, due to the observational biases against detection \citep{Kasliwal2012}. For example: the discovery of the kilonova AT2017gfo \citep{Coulter2017,Kasliwal2017,Arcavi2017,Nicholl2017}, the optical counterpart to the double neutron-star merger event GW170817\citep{Abbott2017,Abbott2017a}, which was a fast, faint, and rare event located in this area of phase space. Other rare classes of transients in this fast part of the phase space have also recently been discovered, such as Fast-Blue Optical Transients \citep[FBOTs; e.g.,][]{Drout2014}, of which the mysterious extreme object AT2018cow \citep{Perley2019} is a likely a member. In fact, \citet{Fox2019} suggest in their recent paper that AT2018cow might be related to Type Ibn SNe.

\begin{figure}
  \centering
  \includegraphics[width=\linewidth]{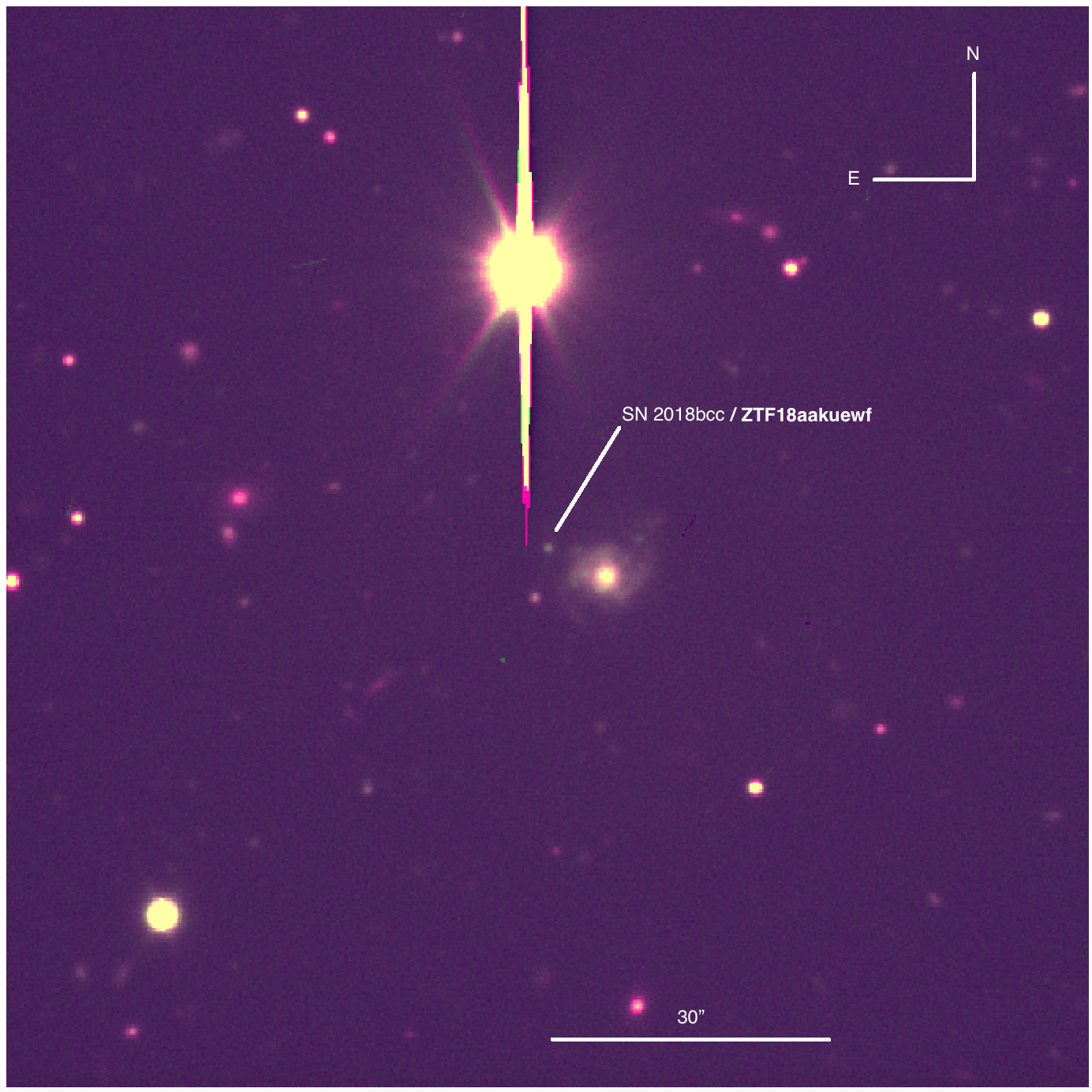}
  \caption{Color-combined image of SN~2018bcc ($g$- and $r$-band filters from TNG+DOLORES) at the epoch of our last detection. North is up and East is to the left. The field-of-view is 2\arcmin x 2\arcmin. The SN, well visible at the center of the image, is located in the outskirts of its host galaxy. A bright, saturated star is positioned to the north and east of the host galaxy.}
  \label{fig:fc}
  \end{figure}

In this paper, we highlight the possibility of using the Zwicky Transient Facility (ZTF), a state-of-the-art wide-field robotic transient survey, to find and follow up on rapidly evolving explosive transients by focusing on the example of SN~2018bcc, which is a fast-rising Type Ibn SN detected during early survey operations (Fig. \ref{fig:fc}). Despite the rarity of Type Ibn SNe, ZTF was able to obtain the hitherto best example of a fast Type Ibn SN with a well-sampled rising LC to date.

The classification of Type Ibn SNe was first proposed by \citet{Pastorello2007} to include H-poor stripped-envelope (SE) SNe with relatively narrow He (${\sim}2000~\text{km}~\text{s}^{-1}$) emission features, with the prototypical example being SN~2006jc \citep{Pastorello2007,Foley2007,Pastorello2008}. Type Ibn SNe are relatively blue and bright at peak compared to an ordinary SE SN \citep{Pastorello2016}. The brightness, blue color, and the relatively narrow He emission features are all thought to be evidence of CSM interaction with a He-rich CSM, which could potentially be the primary powering mechanism of this class \citep[see e.g.,][]{Pastorello2016,Hosseinzadeh2017,Karamehmetoglu2017}.

To date, no more than ${\sim}30$ Type Ibn SNe have been published in the literature \citep{Hosseinzadeh2019}. Most of them have been studied in sample papers by \citet[P16]{Pastorello2016} and \citet[H17]{Hosseinzadeh2017}, which have uncovered several puzzling observational signatures.

H17 show that the class of Type Ibn SNe displays a surprisingly homogeneous photometric evolution for CSM interacting SNe. Nevertheless, in their earlier sample paper, P16 argue that the class shows a large heterogeneity in spectral and LC characteristics, with both very fast and long-lasting examples \citep[see e.g.,][for fast and slow, respectively]{Pastorello2015,Karamehmetoglu2017}. Similarly, a whole gamut of spectral properties can be found in the Type Ibn class that range from those with spectra that look very much like Type Ib SN spectra to those that show a significant number of H emission lines (even though the He features dominate), making them intermediate cases between Type IIn and Type Ibn SNe. Although this heterogeneity was also acknowledged in H17, the Type Ibn SN LCs in their analysis still seem to be much more similar to each other than the LCs of Type IIn SNe. The relative photometric homogeneity can offer clues into the nature of the CSM around Type Ibn SNe. It also means that the detailed study of a well-observed event would prove fruitful.

In another puzzling observational signature, H17 noted that the early \ion{He}{i} line profiles of Type Ibn SN seem to cluster into two groups (primarily based on the strong \ion{He}{i} $\lambda~5876$ emission line): one showing strong P-Cygni profiles and the other appearing primarily in emission. By 20 days past peak however, both groups had evolved to become more emission dominated. H17 suggested that this division could be a result of optical depth effects in the CSM. Since \ion{He}{i} line profiles and fluxes depend on density, temperature, and optical depths \citep[e.g.,][]{Fransson2014}, their suggestion can be investigated by modeling the optical depth effects on the \ion{He}{i} line profiles of Type Ibn spectra.

\begin{figure}
  \centering
  \includegraphics[width=1.05\linewidth]{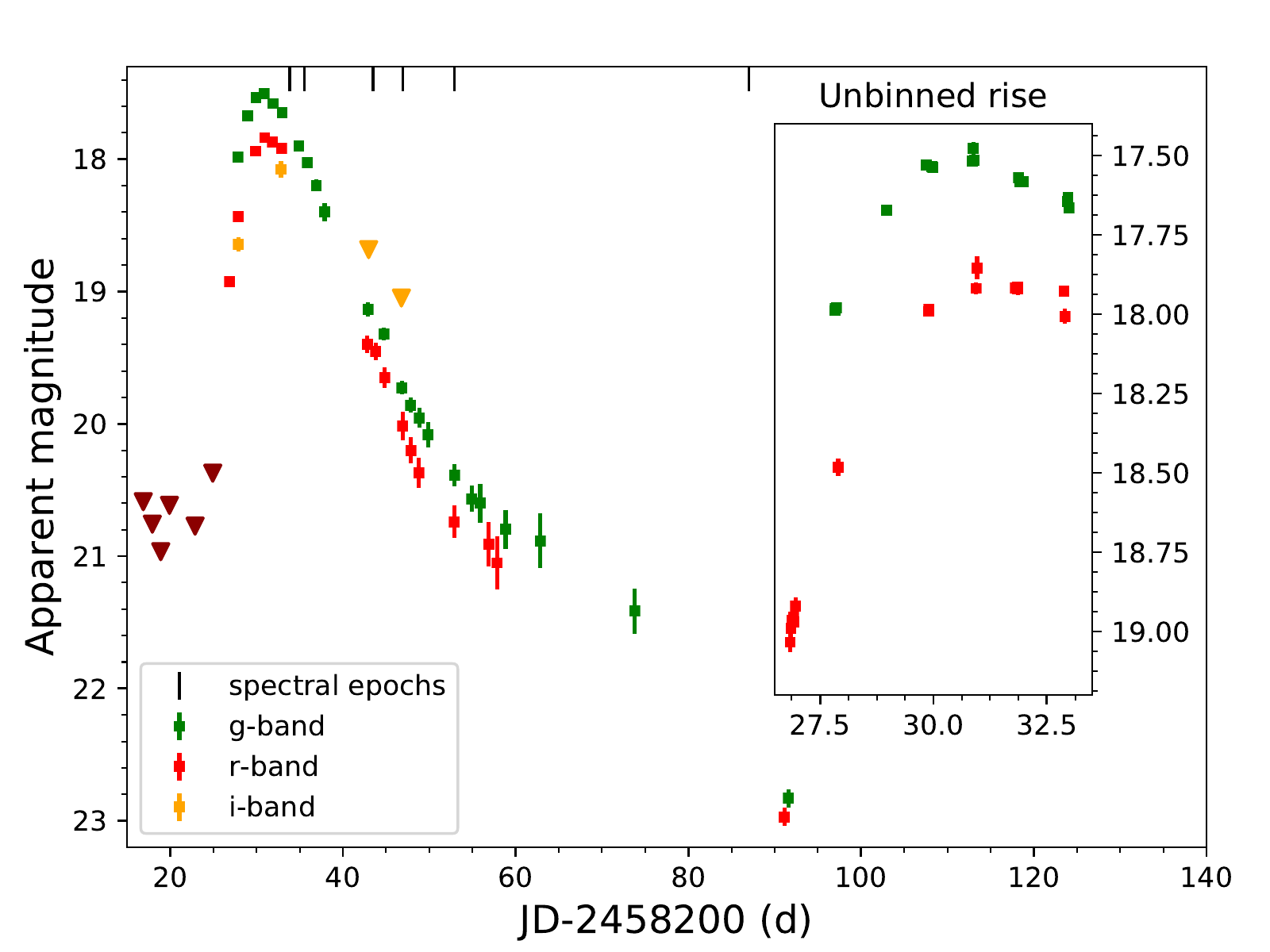}
  \caption{Apparent magnitude LC of SN 2018bcc, binned daily. The average magnitude of each bin was used with the errors added in quadrature. Limits are the deepest limit for the night. The observations are from the P48 survey telescope except for the final $g$ and $r$-band detections, which were obtained with the TNG. The spectral epochs are indicated with vertical black dashes at the top of the figure. \textbf{Inset:} A zoomed-in version of the unbinned rising LC, highlighting the well-captured rise in the $r$ band.}
  \label{fig:appmag}
   \end{figure}

In the rest of this paper, we present the most well-constrained fast-rising Type Ibn SN, SN~2018bcc, and compare it to a selected literature sample. It has a $<6$~d rise time as seen in Fig.~\ref{fig:appmag} and declines on a similar time-scale to other Type Ibn SNe. While several examples of Type Ibn SNe with fast rise times ($<<10$~d) are known in the literature \citep[see e.g.,][H17]{Pastorello2015}, their rapid evolution and the rarity of Type Ibn SNe have resulted in few detections at or before peak. Consequently, observations defining the shape of the rising LC have been lacking for these rapidly evolving events. Additionally, their rapid decline has also made it difficult to obtain late-time spectra.

The paper is laid out as follows: Section~\ref{sec:18bcc} presents basic information about SN~2018bcc and its host. Section \ref{sec:obss} describes the ZTF survey and contains the details of the photometry and spectroscopy that were obtained. The analysis of this data is presented in Sect.~\ref{sec:analysis} followed by our semi-analytical modeling in Sect.~\ref{sec:modeling}. The results are presented in Sect.~\ref{sec:results} and discussed in the context of Type Ibn SNe in Sect.~\ref{sec:discuss}. Finally we list our conclusions in Sect.~\ref{sec:concl}.

\section{SN 2018bcc \label{sec:18bcc}}

SN~2018bcc, located at RA~=~16:14:22.65 and Dec~=~+35:55:04.4 (J2000.0), was first detected by the ZTF on 2018 April 18.34 (JD 2458226.84052) at an $r$-band magnitude of 18.98. Previously, ZTF observed this field 1.89 days before, on April 16.45, and could not detect any source down to a magnitude of $r$ > 20.12 at the location of the transient. In the first three days after discovery, the SN was followed in $r$ band only, and thereafter it was also imaged in $g$ band. In April of 2018, the ZTF survey had begun public survey observations but the associated alert stream \citep{Patterson2019} had not yet started. The transient was subsequently discovered by ATLAS on 2018 April 19, and by Gaia five days later\footnote{\url{https://wis-tns.weizmann.ac.il/object/2018bcc}}.

The Milky Way (MW) extinction towards the SN was estimated to be E(B$-$V)$~= 0.0124$ mag \citep{Schlafly2011}. We did not detect evidence for significant host extinction (in the form of narrow Na ID absorption lines or a reddened spectrum) and assumed it to be negligible for the analysis presented in this paper. Additionally, the SN is located in the outskirts of its host galaxy and is blue in color, which corroborates our assumption of insignificant host extinction. The redshift towards the SN was measured to be $\text{z} = 0.0636$ (Sect.~\ref{sec:specID}) from a fit to the He emission lines of the SN, and verified via host galaxy emission lines present in the latest 2D spectrum. Assuming a cosmology with H$_0=73$~km~s$^{-1}$~Mpc$^{-1}$, $\Omega_\text{M}=0.27$, $\Omega_\Lambda=0.73$ (selected to enable easier literature comparisons), we calculated a distance modulus $=37.19$ mag. This corresponds to a luminosity distance of 274 Mpc.

\section{Observations \label{sec:obss}}

Building on the success of previous transient surveys, the next-generation synoptic surveys such as the Zwicky Transient Facility \citep[ZTF;][]{bellm2019,graham2018}, the successor of the (i)PTF, are now online.

\subsection{The Zwicky Transient Facility (ZTF) \label{sec:ZTF}}
 The ZTF\footnote{\url{http://ztf.caltech.edu}} is a state-of-the-art robotic survey being conducted on the Palomar Samuel Oschin 48-inch (1.2 m) Schmidt Telescope with a 47-deg$^2$ effective field of view. In the standard configuration, the survey takes 30~s exposures reaching down to $\sim$21$^{\text{st}}$ magnitude in the $r$ band with 10~s instrument overheads and a dedicated robotic scheduler that minimizes slewing overheads \citep{Bellm2019a}. Essentially, the ZTF can survey the entire northern sky above Palomar every night down to 21$^{\text{st}}$ magnitude \citep{bellm2019}.

The ZTF survey has been optimized for transient science \citep{Bellm2016} by both covering the entire night sky with a ${\sim}3$~day cadence in $g$ and $r$ bands, whilst also covering smaller parts of the sky with $2\text{--}4$ observations per night in the two filters. As a result, the ZTF survey provides an opportunity to populate some of the remaining gaps in the the brightness and duration phase-space of transients, especially with respect to discovering and obtaining well-sampled LCs for rare and fast objects such as Type Ibn SNe.

\subsection{Photometry \label{sec:phot}}

We collected photometry in $g$, $r$, and $i$ bands from the ZTF survey obtained with the P48 telescope\footnote{The ZTF filter profiles can be found at: \url{https://github.com/ZwickyTransientFacility/ztf_information}}. The ZTF pipeline is described in \citet{Masci2019} and employs the image subtraction algorithm of \citet{Zackay2016}. We also obtained late-time images in $g$ and $r$ bands using the 3.58m Telescopio Nazionale Galileo (TNG). The photometry is made publicly available via WISeREP\footnote{\url{https://wiserep.weizmann.ac.il}} \citep{Yaron2012}.

The P48 photometry was reduced using FPipe \citep{Fremling2016}, which produces point-spread function (PSF) photometry from sky and template subtracted images. The final photometric reduction was done using reference templates obtained by stacking images from more than 2 weeks before discovery (JD~$<2458209$) to remove any possibility of SN light contributing to the template. FPipe photometry is converted to the SDSS photometric system using reference stars in the field selected from the Pan-STARRS 1 survey \citep{Magnier2016,Flewelling2016}. The final epoch of images obtained with the TNG were reduced using standard IRAF\footnote{Image Reduction and Analysis Facility.} photometric reduction steps without template subtraction (since the galaxy is better resolved with the TNG).

SN 2018bcc was detected very early in the ZTF public survey. Some of the data were obtained during verification of subsystems. During normal operations, the ZTF LCs would be more evenly sampled, but here we actually obtained an unusually high sampling during the first night. The photometry for the alerts has since been improved upon, and alert distribution in the public survey began to be published in June 2018. We only work with re-reduced photometry from FPipe in this paper.

  \subsection{Spectroscopy \label{sec:spec}}
\begin{figure}
  \centering
  \includegraphics[width=\linewidth]{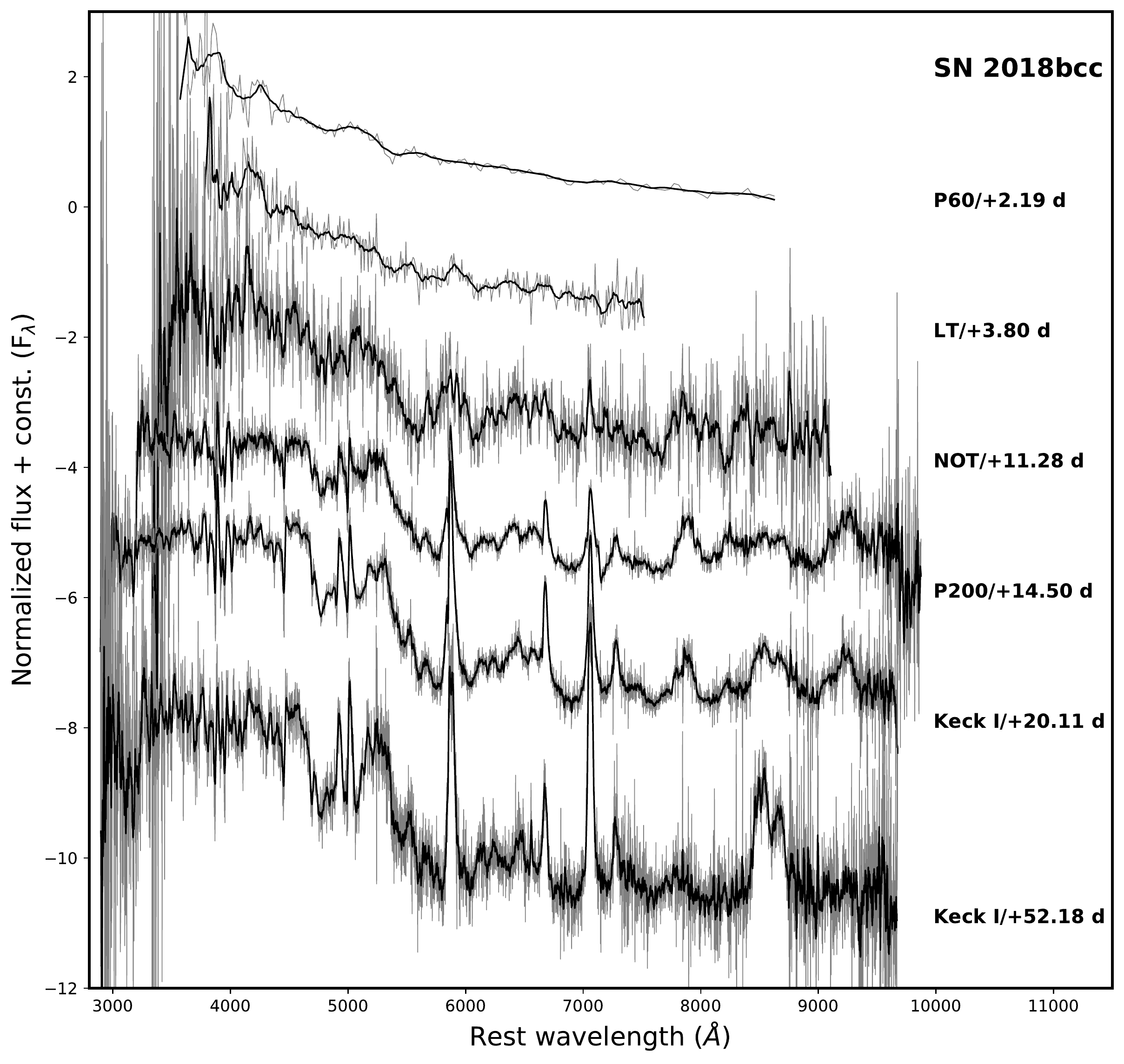}
  \caption{Spectral sequence for SN~2018bcc with restframe phase since peak (JD 2458231.5) of each spectrum indicated. The original spectrum is shown in gray while solid black lines represent the binned data.}
  \label{fig:specseq}
  \end{figure}

A log of the spectral observations is presented in the Appendix, Table~\ref{tab:spectra} and the spectral sequence is show in Fig. \ref{fig:specseq}. An initial spectrum was obtained $6.99$~d after discovery with the Palomar 60 inch (P60) telescope \citep{Cenko2006} equipped with the SED machine \citep[SEDM;][]{Blagorodnova2018,Rigault2019}. Since the SEDM spectrum showed a featureless blue spectrum, an additional classification spectrum was obtained using the SPectrograph for the Rapid Acquisition of Transients (SPRAT) mounted on the Liverpool Telescope \citep[LT;][]{Steele2004}. Relatively low resolution but automated spectroscopic follow-up instruments such as the SEDM and SPRAT are the primary means by which ZTF classifies extragalactic transients \citep{2018ATel11688....1F}.

Additional follow-up spectra were obtained with Keck I equipped with LRIS (Low Resolution Imaging Spectrometer; \citealp{Oke1995,Perley2019a}), the Nordic Optical Telescope (NOT) equipped with ALFOSC (The Alhambra Faint Object Spectrograph and Camera), and the Palomar 200 inch (P200) telescope equipped with DBSP \citep[the Double Spectrograph;][]{Oke1982,Bellm2016b}. The spectra are also made publicly available via WISeREP.

The spectra were reduced by members of the ZTF collaboration in a standard manner with dedicated pipelines for each instrument. In general, the spectra are bias and flat-field corrected, wavelength calibrated using an arc spectrum, extracted and then flux calibrated using a sensitivity function obtained from observing spectrophotometric standard stars. We further absolute flux calibrated the spectra by comparing synthetic photometry computed from the spectra to our $r$-band photometry.

The synthetic photometry were computed using synphot \citep{2018ascl.soft11001S} and the PS1 filter profiles since our photometry were calibrated to PS1\footnote{Obtained from the Spanish Virtual Observatory Filter Profile Service (SVO), http://svo2.cab.inta-csic.es/theory/fps3/index.php}. The spectra were then de-reddened to correct for MW extinction using E(B$-$V)$~= 0.0124$~mag and R$_\textrm{V} = 3.1$, as was the case for the photometry.

\section{Photometric analysis \protect\protect\label{sec:analysis}}
\begin{figure*}
  \centering
  \includegraphics[width=0.8\linewidth]{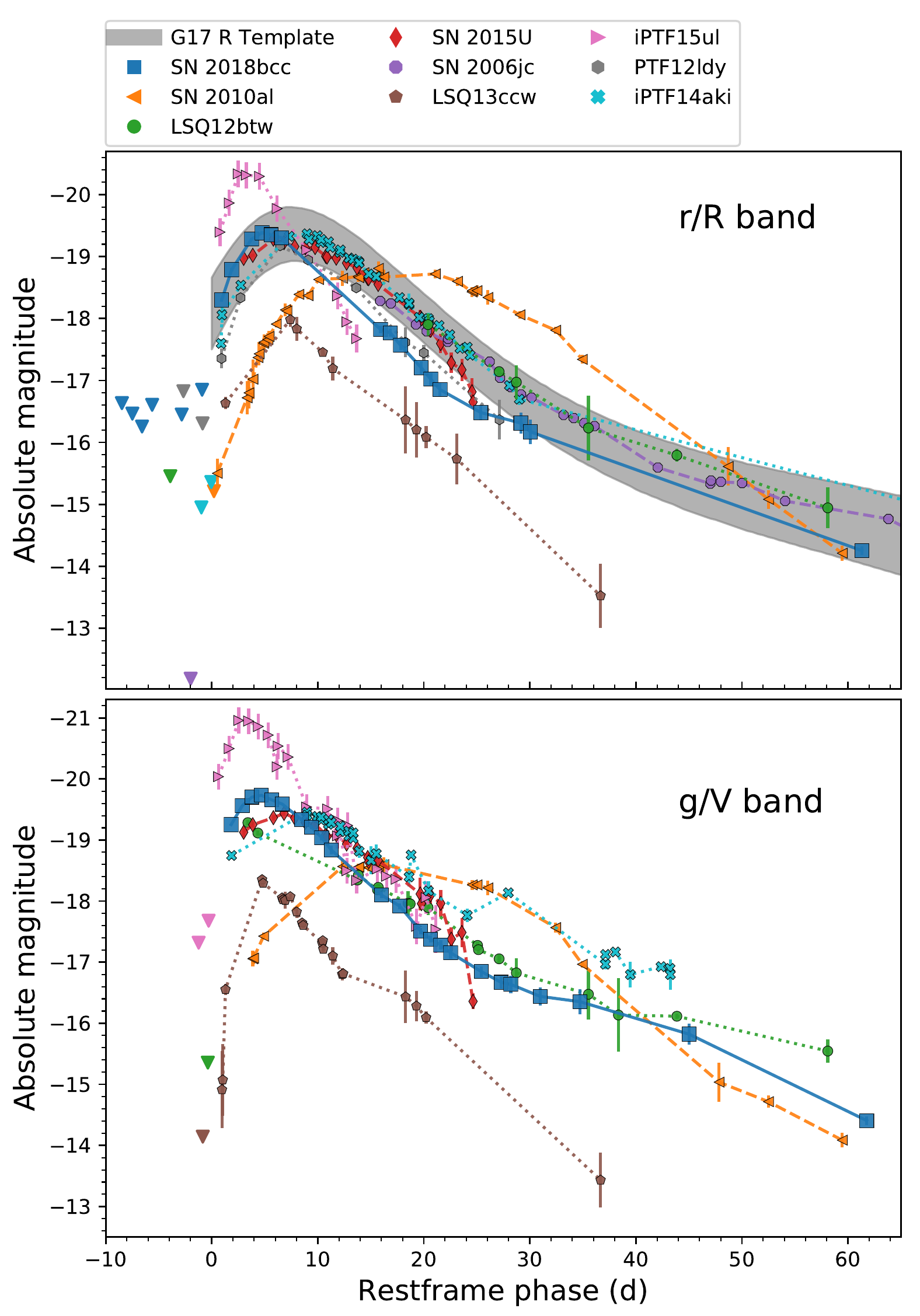}
  \caption{Restframe absolute magnitude LC of SN~2018bbc (binned daily) compared to the literature sample (Sect.~\ref{sec:analysis}). SN~2018bbc is shown with a blue line, the fast-rising SNe are shown with dotted lines, and the other Type Ibn SN LCs are shown with dashed lines. \textbf{The upper panel} shows the $r$/R band while \textbf{the lower panel} contains the $g$/V-band LCs. The LCs are plotted relative to estimated explosion epoch and their respective pre-discovery limits are also shown with triangles. They have been corrected for redshift, MW, and host extinction (where appropriate). The R-band Type Ibn template from \protect\citet{Hosseinzadeh2017} is plotted for comparison, with the phase given since the first available template point.}
  \label{fig:absmag}
  \end{figure*}

We constructed restframe absolute magnitude LCs for SN~2018bcc using the previously mentioned distance modulus. Correction for extinction due to the MW was applied while host extinction was assumed to be negligible (Sect.~\ref{sec:18bcc}). We compare this LC in $g$ and $r$ bands to a literature sample of Type Ibn SN LCs in Fig.~\ref{fig:absmag}. In the comparison sample we included all of the fast rising Type Ibn SNe with a meaningful constraint on the rise: LSQ12btw and LSQ13ccw \citep{Pastorello2015}; PTF12ldy, iPTF14aki, and iPTF15ul \citep{Hosseinzadeh2017}. We also included the well-observed but much slower SN~2010al \citep{Pastorello2015a}, the prototypical example of the Type Ibn class - SN~2006jc \citep{Pastorello2007,Foley2007}, and the very nearby Type Ibn SN 2015U \citep{Shivvers2017} for comparison. The template Type Ibn R-band LC from \citet{Hosseinzadeh2017} is also shown. The literature sample was corrected for MW extinction and we adopted explosion epochs, distance moduli, and redshifts as estimated by the respective works. In the cases of iPTF15ul and SN~2015U, where significant evidence of host extinction is noted and estimated, we also corrected for the host extinction as reported in these papers. We note that without correcting for significant host extinction, iPTF15ul would be similar in brightness to other Type Ibn SNe, including SN~2018bcc.

 \subsection{Rise time \label{sec:risetimes}}
The rise time is the interval between the estimated explosion and peak epochs. Although SN~2018bcc has a fast rise time similar to a few other Type Ibn SNe, the nature of the ZTF observations allowed us to capture an unusually well-sampled rising LC, with seven photometric points on the first day and well-sampled $g,r$ band data on the rise, peak, and decline- the combination of which has not been observed for such a fast-rising Type Ibn SN before (see Fig.~\ref{fig:absmag} for a comparison to other Type Ibn SNe). The well-defined peak epoch and close pre-discovery limits of SN~2018bcc allow us to establish a well-determined rise time; a unique opportunity for a fast Type Ibn SN (Table \ref{tab:risetimes}). We performed this exercise using the $r-$band LC.

\begin{figure}
\includegraphics[width=\linewidth]{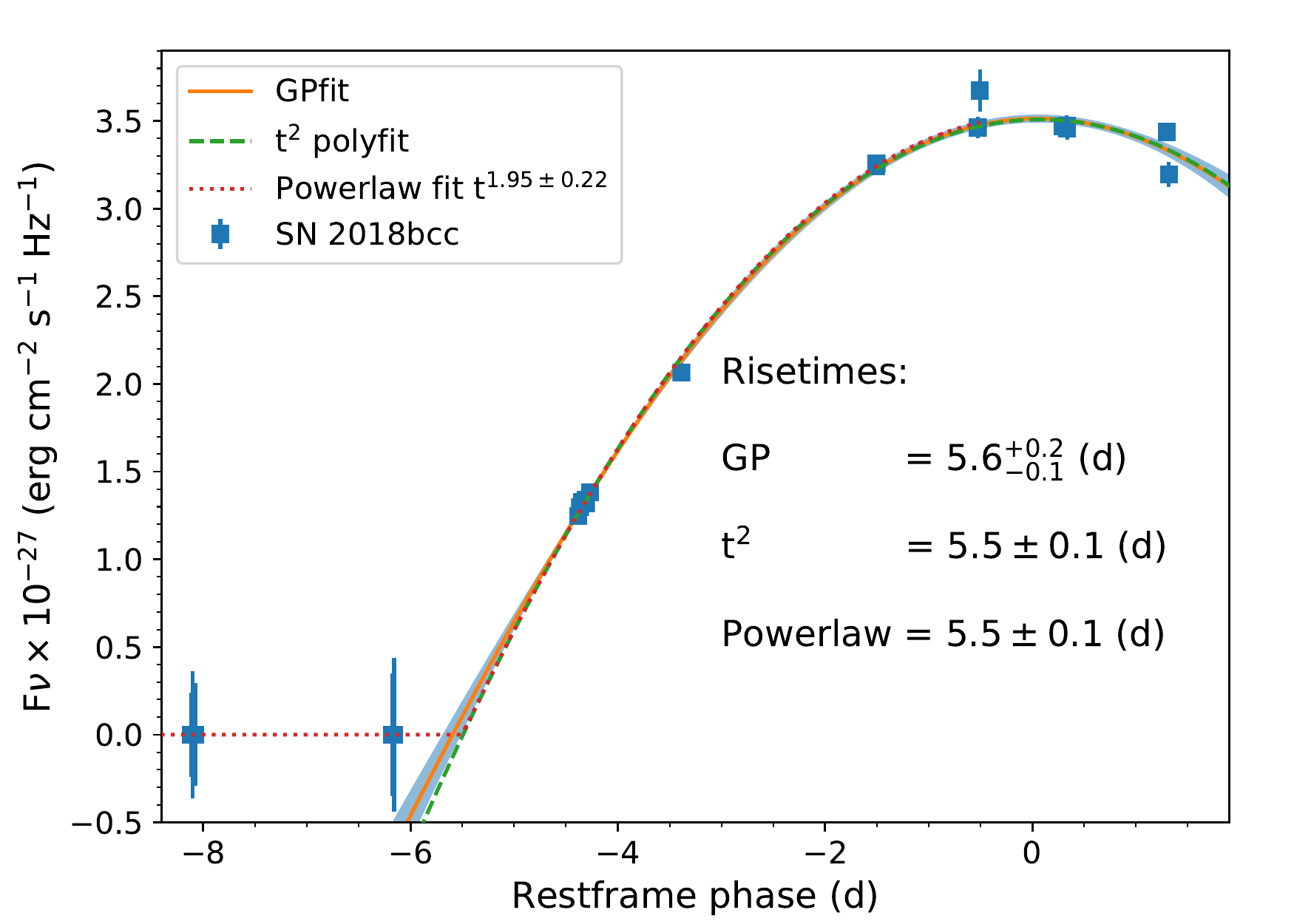}
\caption{Estimating the restframe rise time from fits with $\propto$~t$^2$, a more general t$^{\alpha}$ power law, and using Gaussian process regression, to the rising LC extrapolated to zero flux.}
\label{fig:risetimes}
\end{figure}

\begin{table}
\centering
\begin{tabular}{llll}
\hline
\hline \\[-5pt]
SN & Rise & Uncertainty & Power law $\alpha$ \\
   & (restframe d) & (d) & \\[5pt]
\hline \\[-5pt]
\textbf{SN~2018bcc} &  5.6 & $^{+0.2}_{-0.1}$ & $1.95 \pm 0.22$ \\[5pt]
iPTF15ul  &  4.0 [3.0] & $\pm 0.6~[0.7]$ & $1.8 \pm 0.4$ \\[5pt]
LSQ13ccw  &  [4.7] & $\pm [2.1]$ & - \\[5pt]
LSQ12btw  &  $[<3.8]^\text{a}$ & - & - \\[5pt]
PTF12ldy  &  [6.2] & $\pm [2.0] $ & - \\[5pt]
\hline
\end{tabular}\\[5pt]
{$^\text{\tiny a}$}Peak cannot be determined.
\caption{\label{tab:risetimes} Rise time and power law rising shape of fast Type Ibn SNe, with literature values in square brackets. Calculation of the rise time and the power law fits to obtain the index $\alpha$ for SN~2018bcc and iPTF15ul were performed by us in Sect. \ref{sec:risetimes}. The remaining literature data were taken from the sample paper by \citet{Hosseinzadeh2017}.}
\end{table}

In order to quantify the shape of the rising LC, a low-order polynomial fit was used to obtain the epoch of the peak of the $r-$band LC: JD 2458231.5, which we use as reference phase in the rest of the paper. (The $g-$band peaks slightly before on JD 2458230.9.) The binned LC was put into flux space using the AB-magnitude zero point, with the upper limits shown at zero flux and error bars corresponding to the measured limits. Then, we estimated the explosion epoch and the corresponding rise time (in restframe) in four different ways.

Most conservatively, the rise time was estimated to be the average of the last upper limit and first detection (JD 2458226.8), yielding $5.1 \pm 1.0$~days, with the uncertainty given by the period between last nondetection and first detection. Secondly, we fit the rise in flux space with a $t^2$ scaling polynomial and extrapolated down to zero flux. Assuming that the explosion epoch corresponds approximately to the epoch where the fit crosses zero flux, this method yielded a rise time of $5.5 \pm 0.1$ days. The same assumption is also made in the remaining two methods. Thirdly, we fit a power law of the form {[(t$-$t$_\text{max}$)/t$_\text{rise}\text{]}^{\alpha}$}, while allowing $\alpha$ to vary and ensuring that the upper limits were not violated. We obtained $\alpha = 1.95 \pm 0.22$ and a corresponding rise time of $5.5 \pm 0.1$ days. Finally, we performed a nonparametric fit using Gaussian process (GP) regression, which has the benefit of learning the noise from the data. Extrapolating the GP fit to zero flux yielded a rise time of $5.6^{+0.2}_{-0.1}$ days. The polynomial, power law, and GP fits are plotted in Fig.~\ref{fig:risetimes}. All four estimates of the rise time agree within the errors. For the rest of the paper, we use a rise time of $5.6^{+0.2}_{-0.1}$ days as calculated using the more conservative GP regression, which corresponds to an explosion epoch of JD 2458225.5.

The precise rise time and the shape of the rising LC of SN~2018bcc makes it the most well-observed example among fast Type Ibn SNe. Other fast-rising Type Ibn SNe have uncertain peak epochs due to the difficulty of catching and sampling the peak for such rapidly evolving events. The rise times and rising shapes of fast Type Ibn SNe are compared in Table~\ref{tab:risetimes}. The previously best observed example, iPTF15ul, is the only other fast Type Ibn SN for which we were able to obtain a meaningful constraint on the exponent of the power law fit to the rise, $\alpha = 1.8 \pm 0.4$, and a rise time of $4.0 \pm 0.6$~d using the power law method. The method is the same as described above for SN~2018bcc. The exponents for the power law fits to the two SNe agree within the uncertainties and are close to a value of 2.0. It seems that the large sky coverage and high cadence of ZTF will enable us to obtain a sample of these rare and rapidly evolving events with well-observed peaks and rise times, which has been lacking from previous investigations of fast-evolving Type Ibn SNe.

\begin{figure*}
  \centering
  \includegraphics[width=\linewidth]{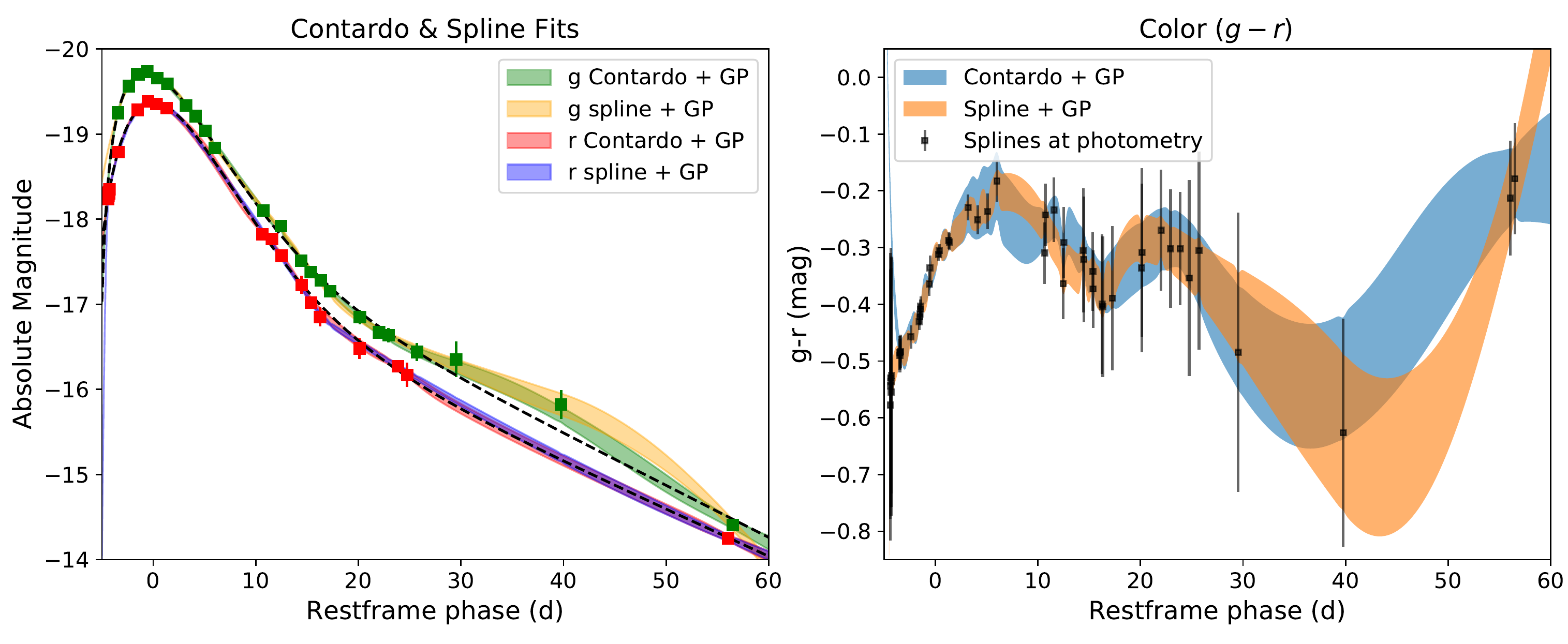}
  \caption{\textbf{Left:} Empirical fits to $g$ and $r-$band LCs used for interpolation. The LCs were fit with both a purely empirical \protect\citet{Contardo2000} model and with smoothing splines. Both these models were modified by a Gaussian process that allowed for both random noise and smooth deviations from the fits. The resulting models were used for interpolation. \textbf{Right:} Restframe $g-r$ color evolution of SN~2018bcc using the interpolations from the left panel. Both methods produced similar results. The black data points represent spline interpolation of the $g$ and $r$ bands to each other at the epochs of photometry with a linear interpolation of the photometric error added in quadrature.}
  \label{fig:colors}
  \end{figure*}

\subsection{Colors \protect\protect\label{sec:color}}
The restframe $g-r$ color evolution of SN~2018bcc is plotted in the right-hand panel of Fig.~\ref{fig:colors}. The colors were obtained by interpolating the $r-$ and $g-$band absolute magnitude LCs using both a \citet{Contardo2000} empirical fit and smoothing splines. Using Gaussian process regression, the empirical fits were assumed to have normally distributed (white) noise around the fit. In addition, the fits were allowed to smoothly vary to better fit the data using a radial basis kernel. This process allows us to build a model that interpolates the full LCs and estimates the uncertainty more robustly, including phases of the LC far away from photometric epochs (which naturally have larger uncertainties). The interpolation models are shown in the left-hand panel of Fig.~\ref{fig:colors} and these were used to calculate the $g-r$ color in the right-hand panel of the same figure.

The $g-r$ color evolved as follows: At early times, it was relatively blue with an evolution to the red that lasted until ${\sim}2$ weeks post discovery. Afterwards, it became bluer again, before ultimately settling to a relatively flat evolution within the estimated uncertainties. This color evolution is typical of Type Ibn SNe, which are usually quite blue at early times \citep{Pastorello2016}.

The results of the interpolation models differ when extrapolated the very early $g$-band LC
due to the assumptions of each model and to the sparsity of early data. However, we find that this does not make a significant difference in our analysis. In addition to color, these interpolation models were used to absolute flux calibrate the spectra to the photometry and for interpolation of the $g-r$ color when constructing the pseudo-bolometric LC (Sect.~\ref{sec:bolo}). Additionally, we calculated the peak absolute magnitude of SN~2018bcc from the interpolation models as $-19.37 \pm 0.03$~mag in the $r$ band and as $-19.68 \pm 0.03$~mag in the $g$ band.

\subsection{Decline rates}
We measured the decline rate parametrized by the magnitude drop within 15 days after peak ($\Delta\text{M}_{15}$), as well as the late-time decline slope ($s$) of SN~2018bcc. For $\Delta\text{M}_{15}$, the interpolation models from the previous section were used, yielding $\Delta\text{M}_{15}$(r)$ =2.25\pm0.01$~mag and $\Delta\text{M}_{15}$(g)$=2.24\pm0.02$~mag. When estimating $s$, we fit a line to the binned LCs, and only considered the points later than 15 and 20 days past peak, thus calculating two separate slope values, $s_{15}$ and $s_{20}$, respectively. We obtained $s_{15}\text{(r)}=0.066 \pm 0.003$~mag~d$^{-1}$ and $s_{15}\text{(g)}=0.072 \pm 0.003$~mag~d$^{-1}$ when cutting before 15 days and $s_{20}\text{(r)}=0.063 \pm 0.001$~mag~d$^{-1}$ and $s_{20}\text{(g)}=0.066 \pm 0.002$~mag~d$^{-1}$ when cutting at 20 days past peak. Similar to other fast Type Ibn SNe, SN~2018bcc thus seems to decline rather rapidly with a late-time rate of ${\sim}0.06\text{–}0.07$~mag~d$^{-1}$ and has a $\Delta\text{M}_{15}$ that is almost twice as large as the typical value for Type Ia SNe.

\section{Spectroscopic analysis \label{sec:specID}}

We determined the redshift from the relatively narrow lines in the supernova spectrum. Using the last Keck spectrum, we measured the five strongest emission lines, and we identify these with strong \ion{He}{i} emission. The deduced redshift was z = $0.06358\pm0.0007$, where the error is the standard deviation from the mean redshift from the five lines. We also checked this with the location of nearby H$\alpha$ in the 2D spectrum of the last Keck spectrum. There was H$\alpha$ from the host nucleus and a nearby \ion{H}{ii} region, as well as slightly blue shifted emission from the SN (which was resolved and not coming from the host, see Sect. \ref{sec:otherlines}.) From this investigation, we obtained a slightly higher uncertainty of $\pm 0.00098$. The larger redshift error corresponds to a typical systematic uncertainty in our velocity measurements of ${\sim}300\kms$.

\begin{figure}
\centering
\includegraphics[width=\linewidth]{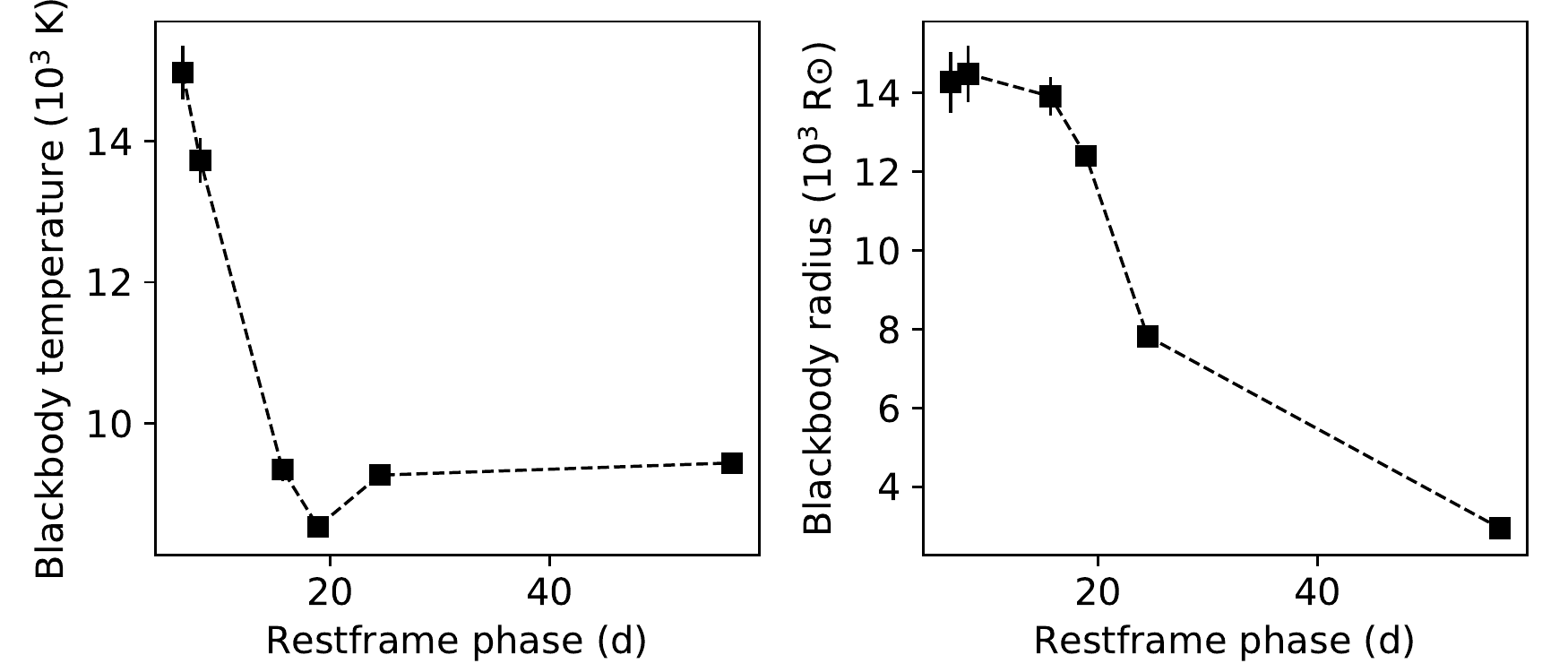}
\caption{\textbf{Left:} Evolution of the blackbody (BB) temperature obtained from BB fits to the spectra. \textbf{Right:} Evolution of the BB radius.}
\label{fig:BBtemps}
\end{figure}

\noindent\begin{figure*}
\centering
\noindent\includegraphics[width=\linewidth]{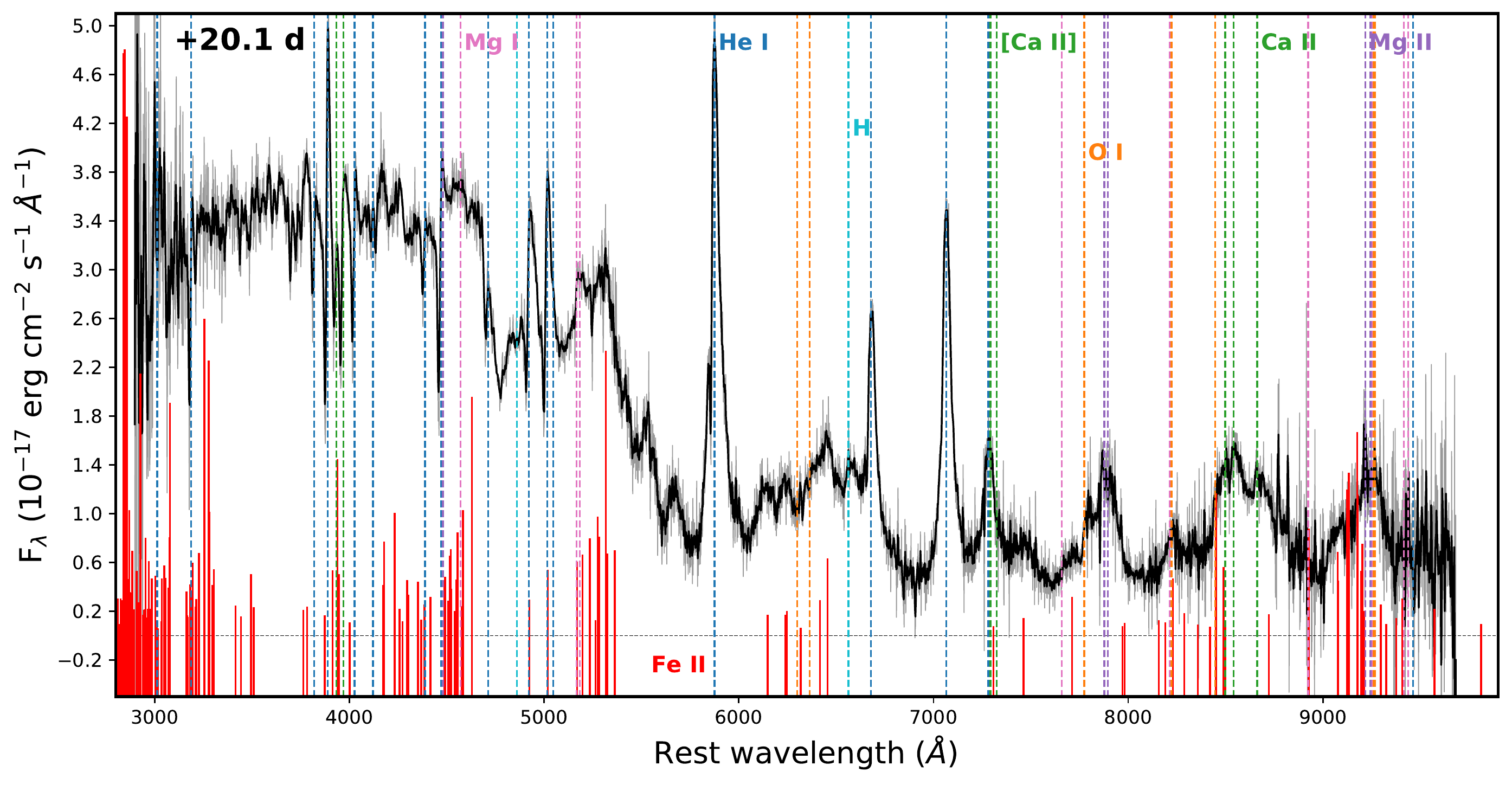}
\noindent\includegraphics[width=\linewidth]{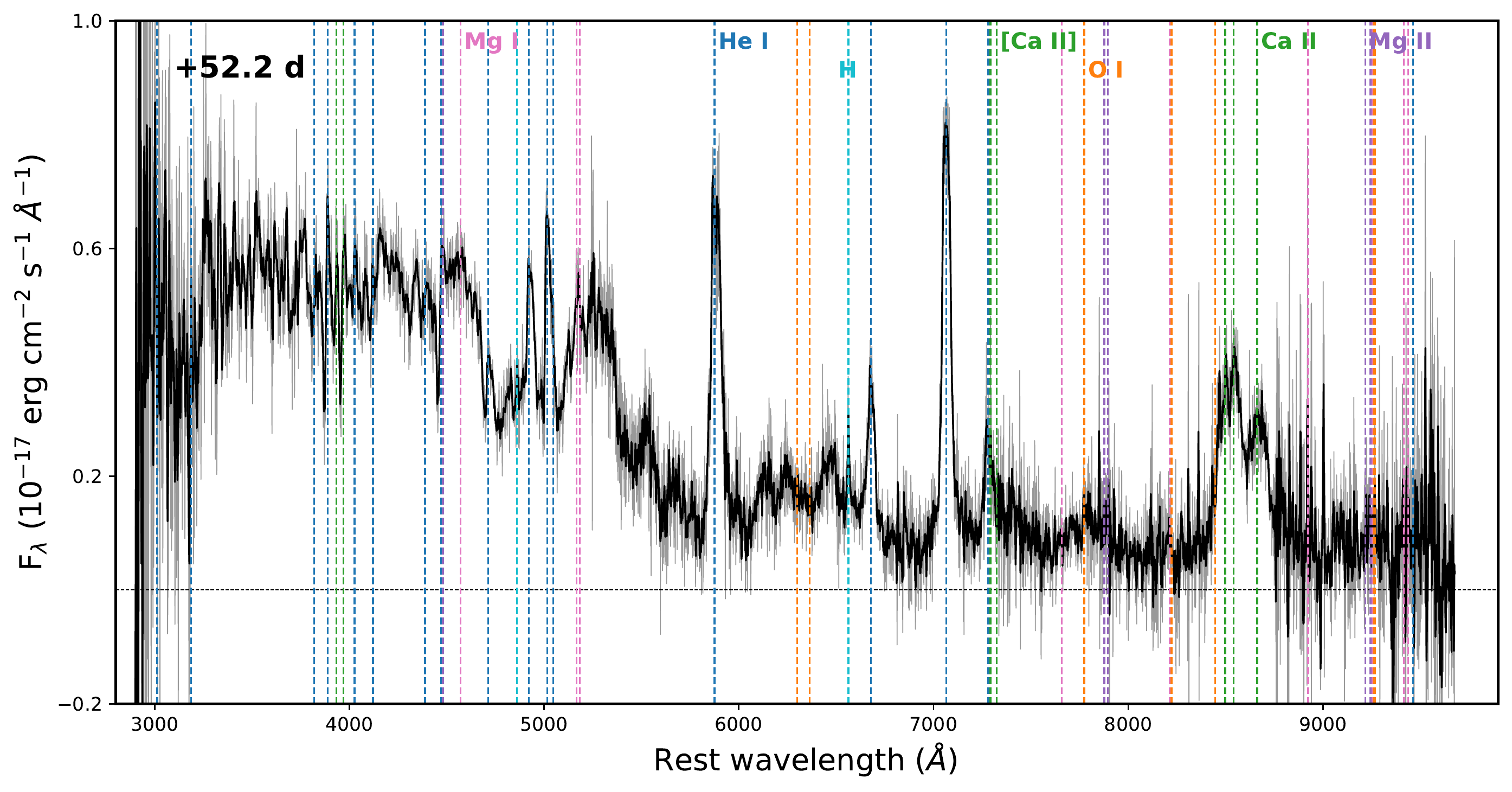}
\caption{\textbf{Upper:} Identification of lines present in spectra. The prominent emission features are marked with dashed lines. Solid red lines show the location of possible \ion{Fe}{ii} features which could be contributing to the spectra. The height of these red lines indicates the expected relative strength of the line. \textbf{Lower:} Line identification carried onto the spectrum at $+52$~d.}
\label{fig:linesID2}
\end{figure*}

Early spectra ($t\lesssim10\,\rm{d}$ past peak) showed a blue featureless continuum, although the signal-to-noise ratio (SNR) was not sufficient to rule out the presence of shallow features. At these early phases, the spectra were well fit by a blackbody with a temperature of $1\text{--}1.5 \times 10^4$~K (see Fig.~\ref{fig:BBtemps} for a plot of the BB temperature evolution). From $+14 \text{d}$, we clearly identified the prominent and relatively narrow \ion{He}{i} emission lines (FWHM $\sim2000$~km~s$^{-1}$), marking SN~2018bcc as a Type Ibn SN. These later spectra are not well matched to a BB due to the presence of strong emission lines, making the effective temperatures obtained from such fits somewhat dubious.

We performed line identifications on the two emission dominated Keck spectra at $+20$ and $+52$ days using common Type Ibn SN lines. The results are plotted in \ref{fig:linesID2}. In addition to the \ion{He}{ii}] lines that dominate the spectra, we also clearly detected lines of \ion{Mg}{i}, \ion{Mg}{ii}, \ion{Ca}{ii}, and late-time H$\alpha$, in addition to possible \ion{Fe}{ii} features (Sect. \ref{sec:otherlines}).

\subsection{Evolution of the \protect\protect\ion{He}{i} lines} \label{sec:Helines}
\ion{He}{i} lines were visible from $+11\,\rm{d}$, although they became more evident at later phases. At $+14\,\rm{d}$, we identified common strong lines of \ion{He}{i}~$\lambda\lambda$
3889, 4471, 5876, 6678, and 7065 in emission, all showing P-Cygni profiles with prominent emission components and a varying strength in absorption. Additionally, we used the strong lines of He tabulated by NIST\footnote{National Institute of Standards and Technology, https://physics.nist.gov/PhysRefData/Handbook/Tables/heliumtable2.htm} and overlaid those with a relative intensity $\geq10$ (where $\lambda~5016$ is 100) on the spectrum taken at $+20\,\rm{d}$, which is plotted with dark blue dashed-lines in Fig.~\ref{fig:linesID2}. As we discuss in Sect.~\ref{sec:Helines}, lab measurements are not appropriate for the conditions in SN~2018bcc, and we only used the line list to guide spectral identification. Nevertheless, every single \ion{He}{i} line from NIST in the wavelength range of our spectrum matched a prominent emission feature with a P-Cygni profile in the spectrum except for in the very edges of the spectrum (\ion{He}{i}~$\lambda\lambda$ 3013 and 9464), where the spectrum was too noisy ($\text{SNR}< 1$). We consider this strong evidence for the detection of 15 \ion{He}{i} lines in the late spectra of SN~2018bcc: $\lambda\lambda$ 3188, 3820, 3889, 4026, 4121, 4388, 4471, 4713, 4922, 5016, 5048, 5876, 6678, 7065, 7281.

\begin{figure*}
\centering
\includegraphics[width=18cm]{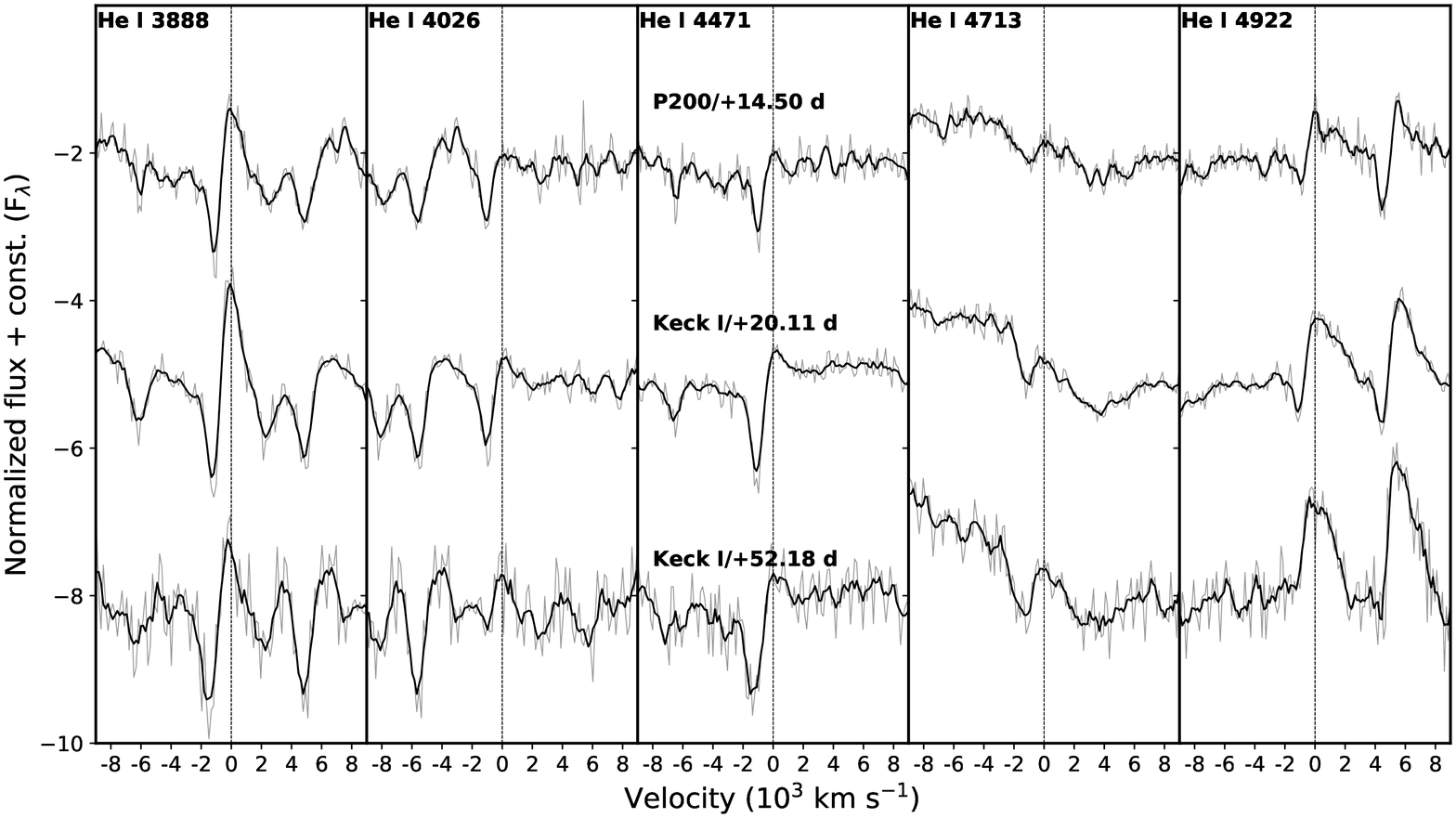}
\includegraphics[width=18cm]{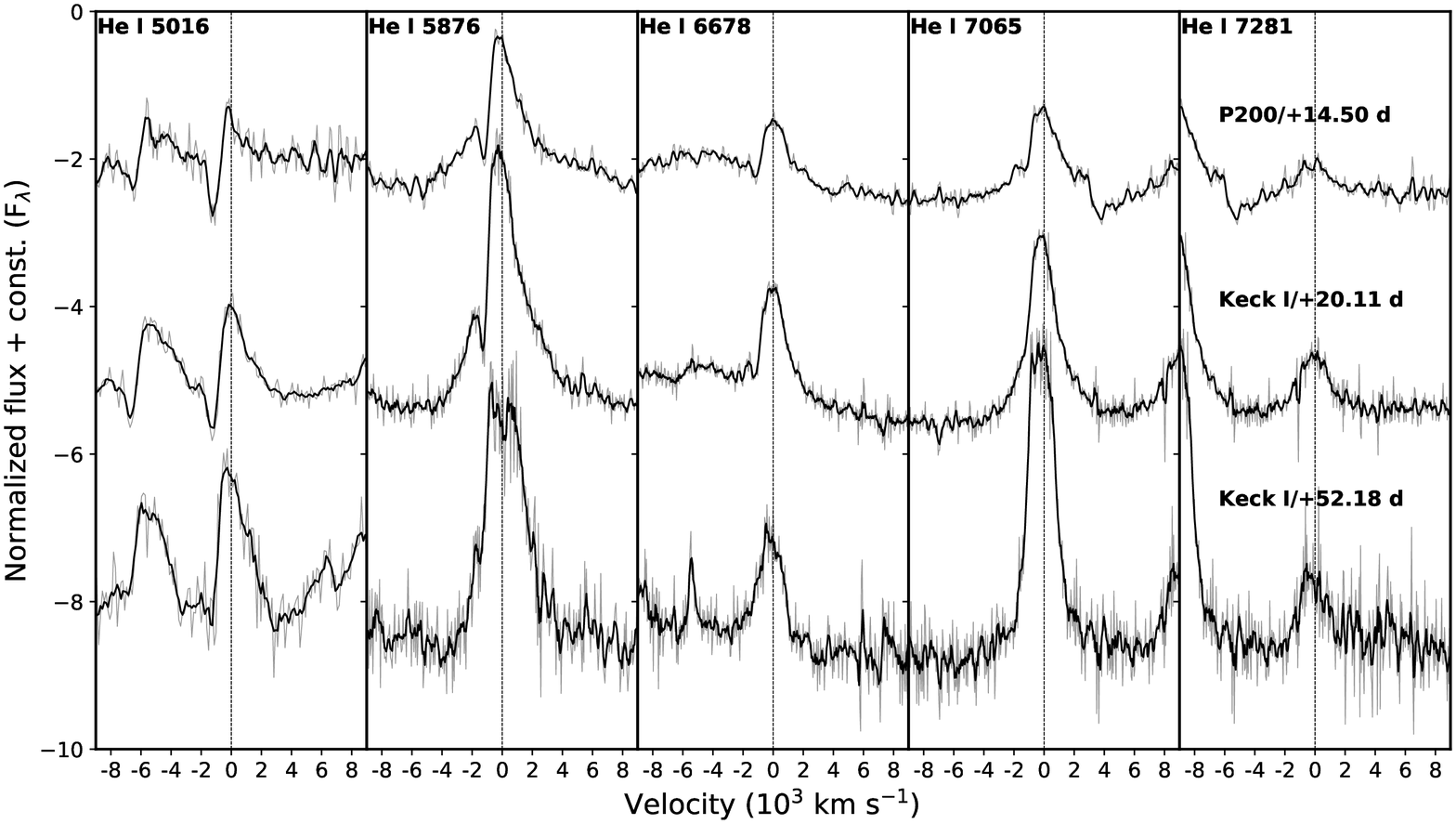}
\caption{Evolution of the ten prominent \ion{He}{i} lines in the latter three spectra. The restframe line locations are indicated with vertical lines.}
\label{fig:HeIlines}
\end{figure*}

All \ion{He}{i} lines showed symmetric profiles with broad wings typical of interacting transients and generally attributed to Thomson scattering by free electrons in an unperturbed dense CSM. The evolution of these lines from +14 to +52~d is shown in
Fig.~\ref{fig:HeIlines}. Fitting a combination of Lorentzian and Gaussian components (for the emission and absorption components, respectively) to the seven strongest \ion{He}{i} observed profiles (Figs. \ref{fig:linesID2} \& \ref{fig:HeIlines}), we inferred FWHMs that are nearly constant with time of $2370\pm530\,\rm{km}\,\rm{s^{-1}}$ for the emission components, while absorption features showed blue wings extending up to $1600\pm200\,\rm{km}\,\rm{s^{-1}}$ and absorption minima of $960\pm100\,\rm{km}\,\rm{s^{-1}}$ (possibly representing expansion velocities).

At $+20$~d, \ion{He}{i}$~\lambda$ 7065 still showed a $\rm{FWHM}\simeq2000\,\rm{km}\,\rm{s^{-1}}$, although the profile was well reproduced using a single Lorentzian component, without clear evidence of any absorption component, even though absorption components were clearly visible in other \ion{He}{i} lines in the same spectrum. At $+52\,\rm{d}$, the \ion{He}{i} lines evolved significantly; whereas \ion{He}{i}$~\lambda\lambda$ 3889 and 4471 lines still showed P-Cygni profiles with pronounced absorptions and weak emission components, the lines at redder wavelengths were instead characterized by structured boxy profiles in emission without a trace of absorption.

In aggregate, the redder \ion{He}{i} lines showed weaker P-Cygni at earlier phases, but then evolved more quickly to show even less P-Cygni and more boxy profiles. Meanwhile, the P-Cygni profiles were stronger for bluer lines in the first spectrum, and while they generally got weaker in the latter spectra, they remained stronger than in the red lines and were still clearly visible at +52~d for the seven bluest lines (Fig. \ref{fig:linesID2}).

\subsection{Identification and evolution of other lines \label{sec:otherlines}}
From +14~d, the spectra showed an excess at wavelengths bluer than $\sim5500$~\AA. These are most likely due to the presence of several \ion{Fe}{ii} multiplets (e.g. multiplets 40, 42, 46). Using the calculation by \citet{Sigut2003}, we identified possible \ion{Fe}{ii} lines and obtained expected approximate line strengths. Although the line strengths are specifically for fluorescence by Ly$\alpha$, this mainly influences the lines in the NIR, and their calculation should give an idea of the transitions that are most important. The expected \ion{Fe}{ii} line strengths are plotted with red lines in Fig.~\ref{fig:linesID2}.

Although we did not unambiguously detect strong and resolved \ion{Fe}{ii} lines, the \ion{Fe}{ii} $\lambda\lambda$ 4924, 5018 lines could be contributing to the \ion{He}{i} $\lambda\lambda$ 4922, 5016, 5048 lines (the latter two are blended). As shown in Table \ref{tab:linesFit}, the FWHM velocities of these specific \ion{He}{i} lines showed a large increase between +14 and +52~d not seen in other \ion{He}{i} lines.

At $+14\,\rm{d}$, two broad features were visible at wavelengths consistent with \ion{[Ca}{ii]}~$\lambda\lambda~7292,\,7324$ (blended with \ion{He}{i} $\lambda$~7281) as well as with \ion{Mg}{ii}~$\lambda\lambda$~7877,\,7896, although here we cannot rule out contamination from \ion{O}{i}~$\lambda$~7774 to the observed feature. From $+20\,\rm{d}$, we also identified the \ion{Ca}{ii} H and K lines and the \ion{Ca}{ii} NIR triplet in emission. The spectrum at $+52\,\rm{d}$ showed significant changes in the line shapes. The \ion{Mg}{ii}~$\lambda\lambda$~7877,\,7896 disappeared into the noise, while the \ion{Ca}{ii} NIR triplet and the blend of \ion{[Ca}{ii]}~$\lambda\lambda~7292,\,7324$ and \ion{He}{i} $\lambda$~7281 became stronger. The evolution of \ion{Ca}{ii} and \ion{Mg}{ii} lines are shown in Fig. \ref{fig:mgca} of the Appendix.

In the final spectrum at $+52$~d, resolved H$\alpha$ that is significantly narrower (and weaker) than the \ion{He}{i} features was clearly seen. The FWHM was $\lesssim 1000$~km~s$^{-1}$ and the centroid was blueshifted by $300$~km~s$^{-1}$. The evolution of the region around H$\alpha$ is also plotted in Fig. \ref{fig:mgca}. The +52~d H$\alpha$ line was clearly resolved as it contains $3-4$ resolution elements. We closely inspected it in the 2D Keck spectrum and found it to be well separated from both a nearby HII region and the galaxy center, both of which also lie along the slit. However, because of the low number of resolution elements, it was difficult to delineate the continuum from the line exactly, and hence to precisely measure the FWHM velocity. We estimated a FWHM velocity of ${\sim}8\text{--}900$~km~s$^{-1}$.

The non-helium lines are especially interesting as indicators of advanced nucleosynthesis. The most interesting diagnostic lines in this respect are the [\ion{O}{i}] $\lambda\lambda~6300, 6364$, and \ion{Mg}{i}] $\lambda~4571$. However, even at day +52 when such indicators should become stronger, \ion{Mg}{i}] was weak and [\ion{O}{i}] was below the noise, if present at all(Fig. \ref{fig:linesID2}).

\subsection{Host-environment properties \label{sec:host}}
 One intriguing aspect of SN 2018bcc is the position within the host galaxy, since the supernova is situated relatively far from the center. The host of SN 2018bcc is SDSS J161422.13+355501.3 and does not have a known redshift. Therefore, we used the redshift measured from the SN spectrum (Sect. \ref{sec:specID}). This galaxy has the following measured SDSS magnitudes:
 $u=19.70\pm 0.07,
 g=18.66\pm 0.01,
 r=18.32\pm 0.01,
 i=18.07\pm 0.01,
 z=17.94\pm 0.04$.

Based on the magnitudes of the host galaxy of SN 2018bcc and the position of this SN with respect to the nucleus, it is possible to estimate the metallicity at the SN position. After correcting for extinction and distance modulus, we used the relation provided by \citet{Sanders2013a} that connects the absolute $g$-band magnitude and the $g-r$ color of the galaxy to its central O3N2 oxygen abundance. We obtained log(O/H)+12=8.39~dex. However, the SN is located in the outskirts of the host, at a projected distance of $7\arcsec$ along the major axis, which corresponds to 1.04 times the radius of the galaxy, R\footnote{Which was obtained from the SDSS.}. Assuming a negative metallicity gradient in the galaxy of $-0.47~\text{dex R}^{-1}$ as in \citet{Taddia2015}, we derived a metallicity [log(O/H)+12] at the position of the SN of ${\sim}7.9$~dex, which is significantly subsolar (Z = 0.16~Z$_\sun$). This estimate would be even lower if there were projection effects.

SDSS J161422.13+355501.3 is also a source in GALEX General Release 6 \citep{Bianchi2013}, with UV fluxes in the far- and near-UV bands (FUV and NUV, respectively). We obtained the fluxes in a $7.\!\!^{\prime\prime}5$ aperture using NED\footnote{The NASA/IPAC Extragalactic Database (NED) is operated by the Jet Propulsion Laboratory, California Institute of Technology, under contract with the National Aeronautics and Space Administration.} and corrected them for interstellar dust extinction following \citet{Salim2007} for the FUV band. We used the extinction corrected FUV band flux to estimate the star formation rate using the calibration from \citet{Salim2007}, and obtained SFR$\sim0.2~\text{M}_\sun~\text{yr}^{-1}$ (for a Salpeter IMF).

\section{Modeling \label{sec:overallmodeling}}
In order to model the possible powering mechanism of SN~2018bcc with semi-analytical LC models, we constructed a bolometric LC from our observations. Due to the fact that we lack multi-band coverage across the EM spectrum (especially in the UV and NIR), we employed a hybrid approach to construct the bolometric LC.

\subsection{Bolometric LC \label{sec:bolo}}

In our method, the spectra are first used to create a quasi-bolometric LC via direct integration of the common integration region from 4000~\AA{} to 8500~\AA{} (we excluded the LT spectrum at +3~d since it did not cover this region). This region is then extended to the UV and NIR by using BB fits, going to the left edge of the U-band in the UV at 3300~\AA. Following \citet[][]{Lyman2013}, the SED for wavelengths shorter than 2000~\AA{} is cut since there should be strong line blanketing there. Instead, the SED is estimated using a linear interpolation from zero flux at 2000~\AA{} to the value of our BB fit at 3300~\AA{} (i.e., the left edge of the typical U-band). We converted the integrated flux from this exercise into bolometric magnitudes and calculated a bolometric correction to the $r$-band absolute magnitude LC at the epochs of the spectra using the interpolated $r$-band LC model from Sect. \ref{sec:color}. The bolometric LC was then created from these bolometric corrections. Details of our methods are explained in Appendix \ref{app:bololc}.

The derived bolometric correction is good enough for our purposes; we mainly want to use this bolometric LC to assess the underlying powering mechanisms. The pseudo-bolometric LC we obtained by applying our derived bolometric correction to the $r$-band LC is plotted in Fig.~\ref{fig:bolo}. Except for the systematic uncertainties on the distance and MW extinction, all other sources of error were propagated for this figure, with the uncertainty from absolute flux-calibration being dominant. In the proceeding analysis, we will add the uncertainty from distance (a large systematic error due to cosmology). We also add the scatter in \emph{all} possible bolometric corrections obtained via varying our assumptions (see Appendix \ref{app:bololc}). Thus, the resulting uncertainty is very conservative. While neither error dominated, the scatter in the bolometric correction was a larger source of error. With these additions, the bolometric luminosity at peak ($L_p$) was $2.0 \pm 0.8(0.1) \times 10^{43}$~erg~s$^{-1}$ and the total radiated energy was $2.1 \pm 0.8(0.1) \times 10^{49}$~erg. The error bars in the parenthesis were obtained from the single method described above (and plotted in Fig.~\ref{fig:bolo}), without the larger error from distance and the scatter in all possible values of the bolometric corrections.

\begin{figure}
\centering
\includegraphics[width=1.05\linewidth]{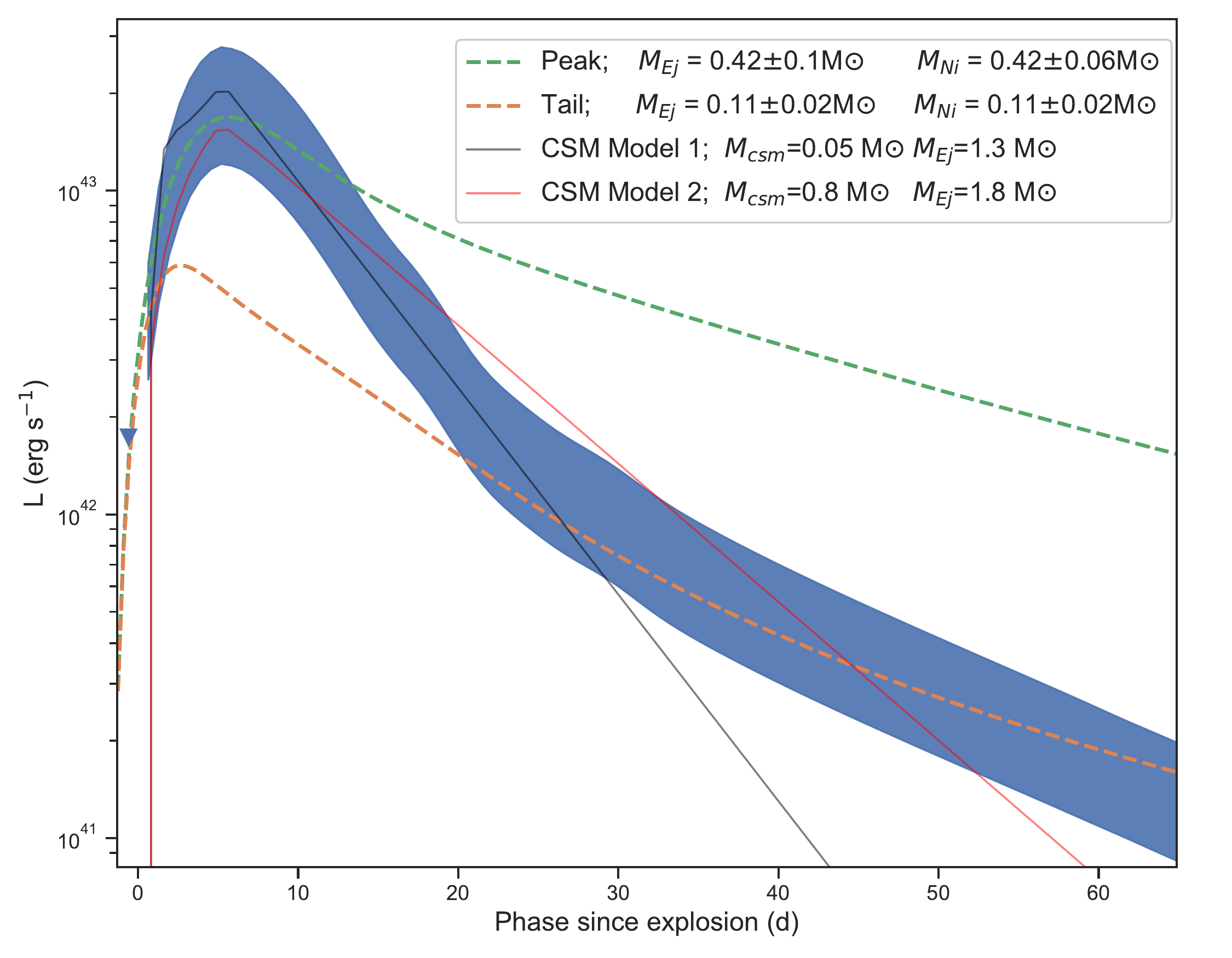}
\caption{Arnett (dashed-lines) and CSM model (straight lines) fits to the bolometric LC of SN~2018bcc (blue). Since the Arnett models cannot fit both the bright peak and the rapidly declining tail simultaneously, even when accounting for gamma-ray leakage, we also show two examples of CSM-interaction powered model fits that can reproduce the bright and rapidly evolving peak of SN~2018bcc; the latter models use different parameters describing the CSM. Due to the caveats of these models (see text), we only use them to illustrate that bright and rapidly evolving LCs, such as that of SN~2018bcc, can be naturally produced by CSM interaction. The r-band upper limit has been included using the same bolometric correction as in Sect. \ref{sec:bolo}.}
\label{fig:arnett}
\end{figure}

\subsection{The powering mechanism of SN~2018bcc \label{sec:modeling}}

While the luminosities of ordinary SE~SNe are thought to be powered by the radioactive decay of \element[][56]{Ni}, strongly CSM-interacting SNe could be dominated by the luminosity input from shocks created by the collisions between the ejecta and the nearby dense CSM. In the case of Type Ibn SNe, it is certainly possible that CSM interaction dominates the luminosity input, since we see evidence for CSM interaction in the spectra. In this section, we model the bolometric LC of SN~2018bcc using two semi-analytical LC models.

In the first, sometimes referred to as a "simple" \citet{Arnett1982} model, we assume the SN is powered only by the radioactive decay of \element[][56]{Ni}. A working algorithm for this model is presented by \citet{Cano2013}. We followed the assumptions of \citet{Karamehmetoglu2017} and used an ejecta expansion velocity of V$=7000$~km~s$^{-1}$, a constant effective opacity of $\kappa = 0.07 ~\text{cm}^2~\text{g}^{-1}$, and an $\text{E}/\text{M}_\text{ej} = (3/10)\text{V}^2$, where E is the kinetic energy and M$_\text{ej}$ is the ejecta mass. In the standard Arnett model, the energy produced by radioactive decay is primarily radiated in the form of gamma-rays that are fully trapped by the ejecta. However, as the opacity drops with expansion, some gamma-rays should escape without thermalizing, thus lowering the observed luminosity. In our modeling, we also took into account the luminosity decrease due to this gamma-ray leakage following the scaling relations of \citet{Clocchiatti1997} and additionally applied the modification for energy loss via positrons \citep{Sollerman1998}, using the same implementation as in \citet{Karamehmetoglu2017}.

The result of our fitting is plotted with dashed lines in Fig. \ref{fig:arnett}. Using these assumptions, it was not possible to explain the rapidly evolving LC of SN~2018bcc using radioactive powering alone. Since there cannot be more nickel than the total ejecta mass, attempting to fit the peak requires a high mass of \element[][56]{Ni} and therefore too high of an ejecta mass to fit the tail of the LC. Conversely, lowering the ejecta mass to fit the tail means that even if all of the ejecta were made of \element[][56]{Ni}, there is not enough to power the peak of the LC. (A more realistic ratio of \element[][56]{Ni} to ejecta mass would allow ${\sim} 2\Msun$ of ejecta that can still fit the tail for the same \element[][56]{Ni} mass.) Thus, it was not possible to get a good fit to the LC of SN~2018bcc with radioactive decay alone.

In order to unconstrain the ejecta mass as much as possible, we allowed the explosion epoch for the Arnett fits in Fig.~\ref{fig:arnett} to vary as long as they were below the pre-explosion limit from the $r$ band, since uncertainty in the explosion epoch is the strongest source of error for ejecta mass. The assumed characteristic ejecta velocity of V$=7000$~km~s$^{-1}$, which was obtained for the transitional Type Ibn SN~2010al by \citet{Pastorello2015a}, is another source of uncertainty. Higher velocities would lead to higher ejecta masses. However, higher velocities also lead to greatly increased gamma-ray leakage \citep[e.g.,][]{Sollerman1998,Karamehmetoglu2017}. These two effects can combine to produce a reasonable fit to SN~2018bcc with $\text{M}_\text{ej} \gtrsim 2\times$ \element[][56]{Ni} mass, but such a model required ejecta velocities in excess of $0.1c$. Since we lack any evidence to support such high-energy ejecta with relativistic velocities, this scenario seems unlikely.

Since it was not possible to fit the LC with a purely radioactive decay-powering scenario using reasonable, observationally based parameters, we turned to CSM interaction as an additional powering scenario that might be able to explain the bright but rapidly evolving LC. In this exercise, we adopted the models of \citet{Chatz2012} to construct a toy model of CSM interaction to test whether CSM interaction can indeed power the bright and rapidly evolving LC of SN~2018bcc. In this simplified model the luminosity from a forward and reverse shock due to ejecta-CSM interaction is assumed to come from a steady photosphere at some suitable distance. We used the same physically based assumptions as in \citet{Karamehmetoglu2017} to construct a CSM interaction model that can fit Type Ibn SNe.

The complicated nature of CSM interaction means that either changing the parameters in the same model, or modifying the model to take more parameters into account, are possible ways to achieve a model fit. To illustrate this fact we plot two example CSM model fits using different parameter values in Fig. \ref{fig:arnett} that both produce a bright and rapidly evolving peak. By showing two different example fits, we highlight that the parameters used are not unique.

The CSM models 1 and 2 had reasonable ejecta and CSM masses of 1.3 and 0.05 $\Msun$ and 1.8 and 0.8 $\Msun$, respectively, with additional differences in parameters controlling the CSM properties (their details can be found in the Appendix). We emphasize that modifying the parameters that control the nature of CSM interaction could extend the model to also fit the tail. In this exercise we simply wanted to show that, unlike in the radioactive decay powered scenario, a bright and rapidly evolving peak can be powered by CSM interaction using reasonable values.

Presumably, if CSM interaction powers the main peak of the LC, an ordinary SE~SN LC with a low \element[][56]{Ni} mass could be hidden underneath. Subtracting the CSM-powered LC could then provide an estimate of the \element[][56]{Ni} mass of the underlying "hidden" LC. However, since our spectra showed continued evidence of CSM interaction and no evidence of SN ejecta, in the form of broad SN lines or a nebular spectrum with enhanced abundances, we caution that any such combined model would have to predict continued CSM interaction and lack spectral features of ejecta from an SN explosion. Nevertheless, using our CSM models from Fig. \ref{fig:arnett}, we estimated that ${\sim} 0.03 \Msun$ of \element[][56]{Ni} could power a possible "hidden" SE-SN LC. A higher \element[][56]{Ni} mass is possible based on the Arnett model fit to the tail, if the CSM interaction is less luminous than our models.

The shock-interaction models of \citet{Chatz2012} were developed for studying SLSNe with potentially high-mass and extended CSM. The assumptions of their simplified models may not be particularly appropriate for rapidly evolving lower-mass objects, such as Type Ibn SNe. Therefore, in order to move beyond our simple modeling, we make use of our spectra to model the properties of the CSM around SN~2018bcc in the next sections.

\subsection{Modeling the \ion{He}{i} line profiles \protect\protect\label{sec:HeIFlux}}

The spectra of SN~2018bcc also offer insights into the nature of the CSM interaction. An important indicator of strong CSM interaction is the presence of a broad electron scattering wing in the emission lines. As can be seen in Fig. \ref{fig:HeIlines}, these scattering wings were clearly detected in the strong \ion{He}{i} $\lambda~5876$ and $\lambda~7065$ emission lines up to day 20, where the wings extended to $\sim 5000 \kms$. Overall the lines were symmetric; a characteristic of electron scattering. As a result, the maximum extension of the lines are likely coming from electron scattering instead of representing the expansion velocity of the emitting region. However, the slightly blueshifted peaks also seen in Fig. \ref{fig:linesID2} indicate that, in addition to electron scattering, there is macroscopic bulk motion of the emitting region. Although, redshift uncertainty can possibly account for some of the observed blueshift (${\sim}300\kms$; Sect. \ref{sec:spec}), which we consider in the following investigation.

\begin{figure}
\centering
\includegraphics[width=\linewidth]{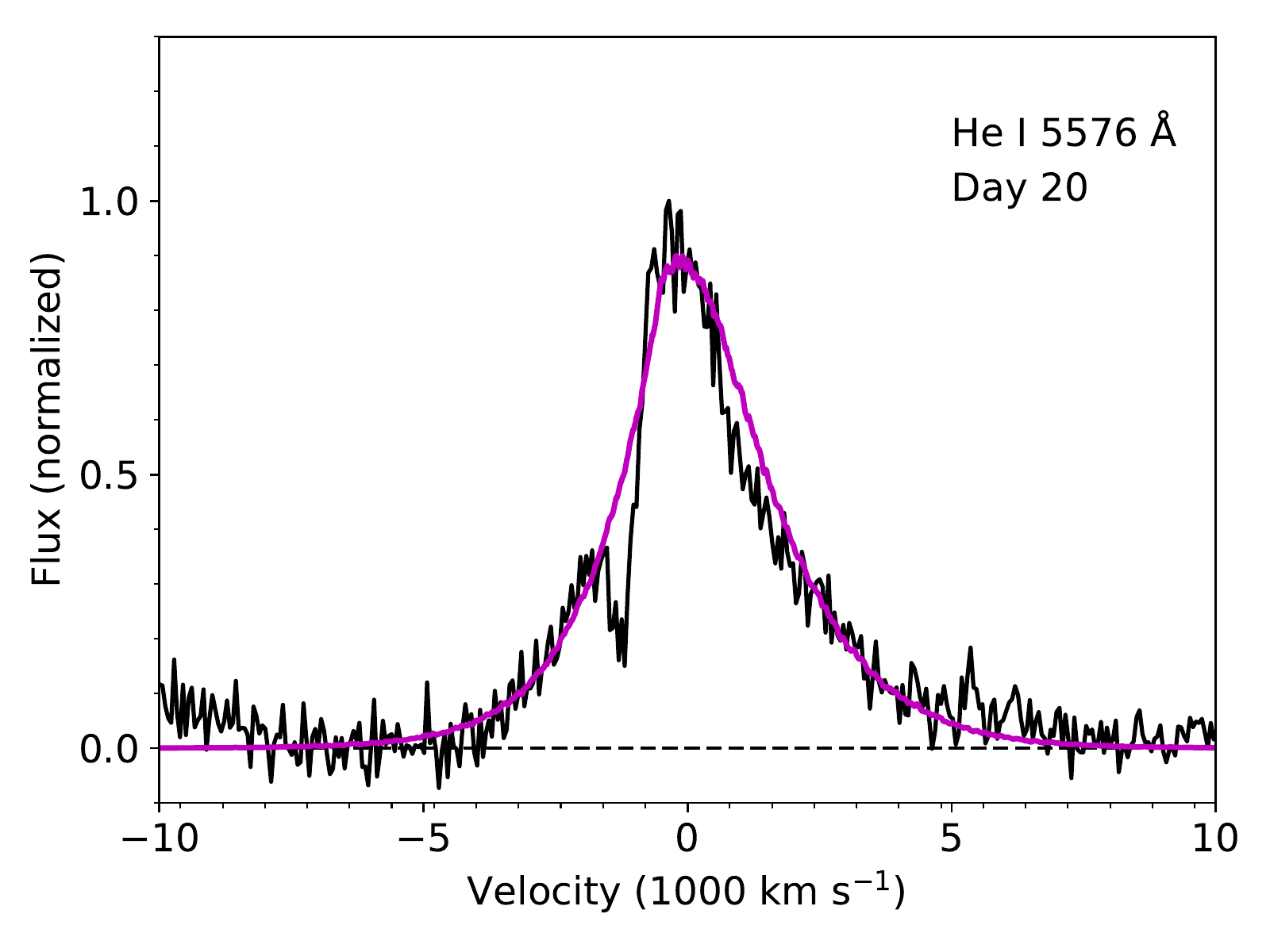}
\caption{Comparison of observed continuum subtracted  \protect\ion{He}{i} $\lambda~5876$ line at day 20 (black) and the calculated line profile due to electron scattering (magenta).}
\label{fig:MCsimul}
\end{figure}

To illustrate this more quantitatively, we modeled the line profile of the day +20 \ion{He}{i} $\lambda~5876$ line, using an extended version of the Monte Carlo code in \cite{Fransson2014}. The result of our modeling is shown in Fig. \ref{fig:MCsimul}, where we have over-plotted our model on top of the continuum subtracted \ion{He}{i} $\lambda~5876$ line. We obtained an expansion speed of the emitting region of $600~\pm 300~\kms$ and an electron scattering optical depth of $\tau_{\rm e}=7$ for an assumed electron temperature of $T_{\rm e} = 15,000$ K. Depending on the assumed temperature, the optical depth scales as $\tau_{\rm e} \propto T_{\rm e}^{-1/2}$. A temperature of 7000 K would therefore correspond to $\tau_{\rm e} \approx 10$.

Given the small number of parameters in the model, the fit is very good. Both the overall shape of the line wings and the blueshift of the peak are well reproduced. The main discrepancy is the P-Cygni absorption between ${\sim}800\text{--}1800 \kms$.

The fact that the electron scattering optical depth is large means that any absorption component  will be filled in by the scattering as long as the optical depth is $\tau_{\rm e} \gtrsim 1$. The P-Cygni lines we observed must therefore form at lower $\tau_{\rm e}$, i.e., in the outer part of the shell, while most of the emission, seen in the $\lambda\lambda~5876, 6678, 7065, 7291$ lines, should originate at larger $\tau_{\rm e}$.

The best estimate of the expansion velocity of the line forming region comes from the blue edge of the P-Cygni lines, in particular the lines with a weak emission component, like the \ion{He}{i} $\lambda~3889$ and $\lambda~4471$ lines. On the day 20 spectrum, these lines extended to $1900 \pm 400 \kms$, which we have taken as the expansion velocity of the CSM. The error was derived by adding in quadrature the measurement uncertainty with the systematic uncertainty from redshift.

The fact that the velocity from the P-Cygni absorption ($1900 \pm 400 \kms$) was larger than that inferred from the line shift of the electron scattering ($\sim 600 \pm 300 \kms$) means that the emitting region of the photons has a lower velocity than where the P-Cygni absorption arises. Such a case is seen in Eta Carinae, where several velocity components from different parts of the ejecta are present. While most of the ejected material has a velocity of $\sim 650 \kms$, a small fraction has velocities up to $3500-6000 \kms$ \citep{Smith2008}. These may be the result of several separate eruptions with different velocities, as observed, for example, for the pulsational pair instability models in \cite{Woosley2017}. However, a range of velocities can also be a result of a single ejection, where a nearly homologous velocity field is achieved after one or two expansion timescales. This type of velocity field is again observed for the ejecta in Eta Carinae \citep{Smith2006}.

Moreover, multicomponent velocity fields have previously been observed in Type Ibn SNe and are discussed by \citet{Pastorello2016}. As they note, the velocity of the emitting material might not represent the velocity of the deepest ejecta layer. Instead, it could also be coming from the interface between the forward and reverse shocks, effectively hiding the ejecta velocity.

\subsection{CSM optical depth and density from \ion{He}{i} emissivity \label{sec:emiss}}
The relative fluxes of the \ion{He}{i} lines are of interest for understanding the physical conditions in the CSM. The strongest lines in the optical region are the $\lambda~5876$ and $\lambda~7065$ lines. Under normal ISM conditions the \ion{He}{i} $\lambda\lambda~7065/5876$ ratio is $\sim 0.18$ \cite[e.g.,][]{Benjamin2002}. However, in SN 2018bcc this ratio was $\sim 0.60$ at +20 days and $\sim 1.01$ at +52 days. The \ion{He}{i} line ratios depend on density, temperature, and optical depth. In particular, a high optical depth of the $\lambda~3889$ line will increase the $\lambda~7065$ line relative to the $\lambda~5876$ line \cite[e.g.,][]{2006agna.book.....O}. Although, other lines are also sensitive to optical depth effects.

To study these effects, we have calculated line strengths for different optical depths, electron densities, and temperatures, using atomic data from \cite{Benjamin2002}. The temperature and electron densities were assumed to be constant in the emitting region and are mainly important for the role of collisional excitation and de-excitation. As is common, we use the optical depth of the $\lambda~3889$ as a parameter. The model assumes a Sobolev escape probability with homologous expansion velocity. As was discussed in last section, this is a natural choice if the CSM is a result of discrete eruptions. However, as long as the lines are optically thick, which is the case considered here, the escape probability is only weakly sensitive to the velocity field. For example, if we consider a constant velocity rather than a homologous expansion, the escape probability is $\beta=2/(3 \tau)$ compared to $\beta=1/\tau$ in the optically thick limit. The factor $2/3$ comes from the an gle averaging.

\begin{figure*}
\centering
\includegraphics[width=9cm]{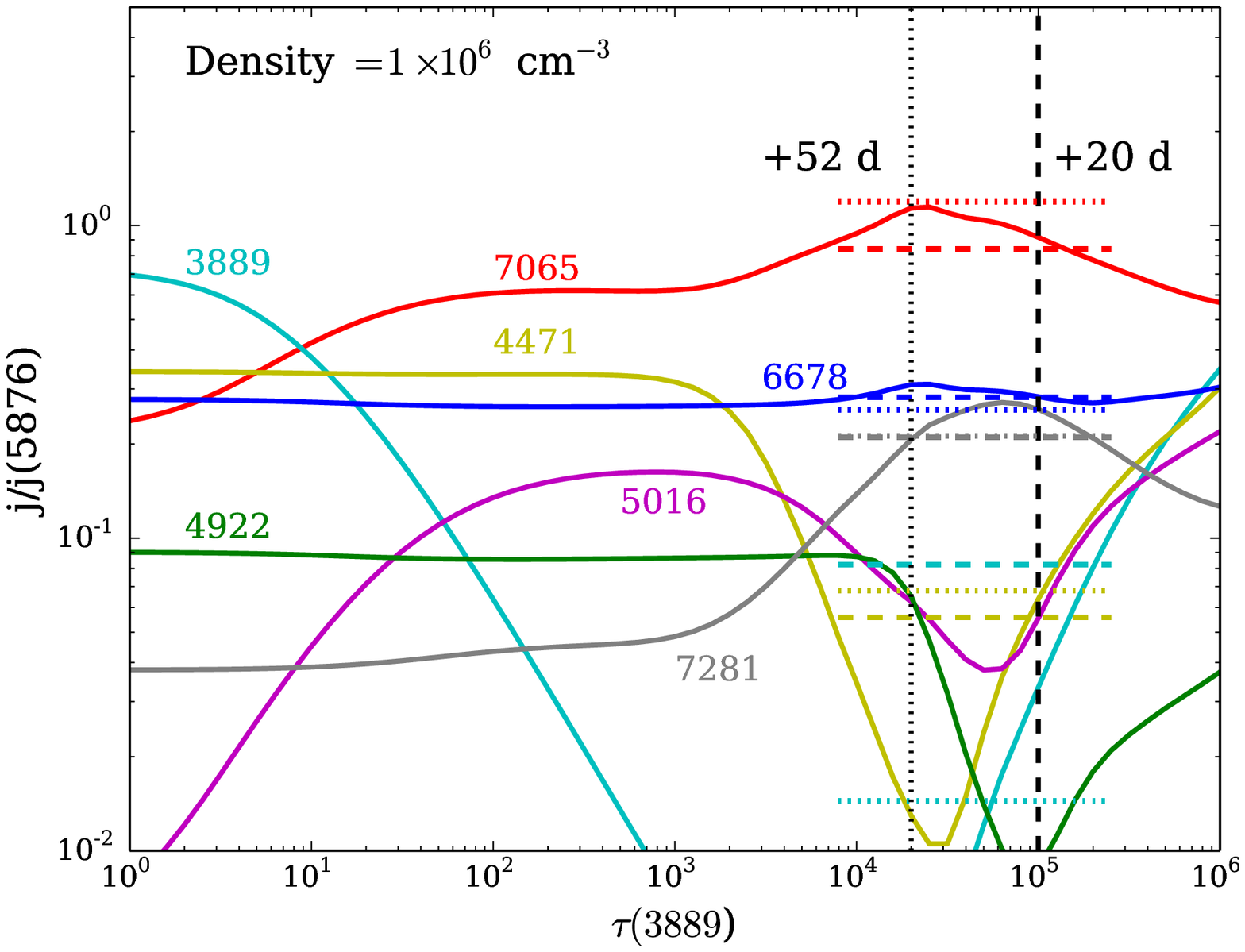}
\includegraphics[width=9cm]{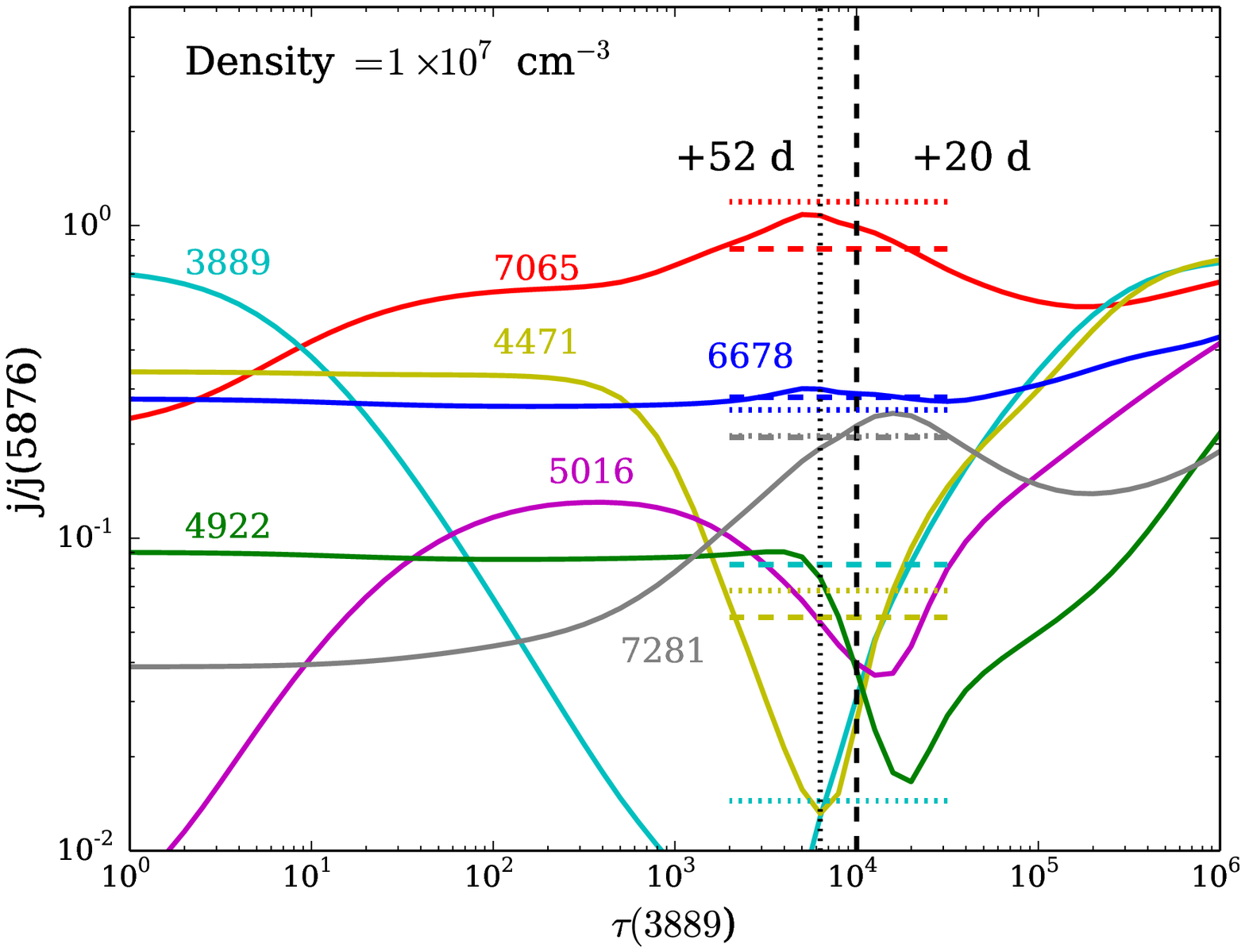}
\includegraphics[width=9cm]{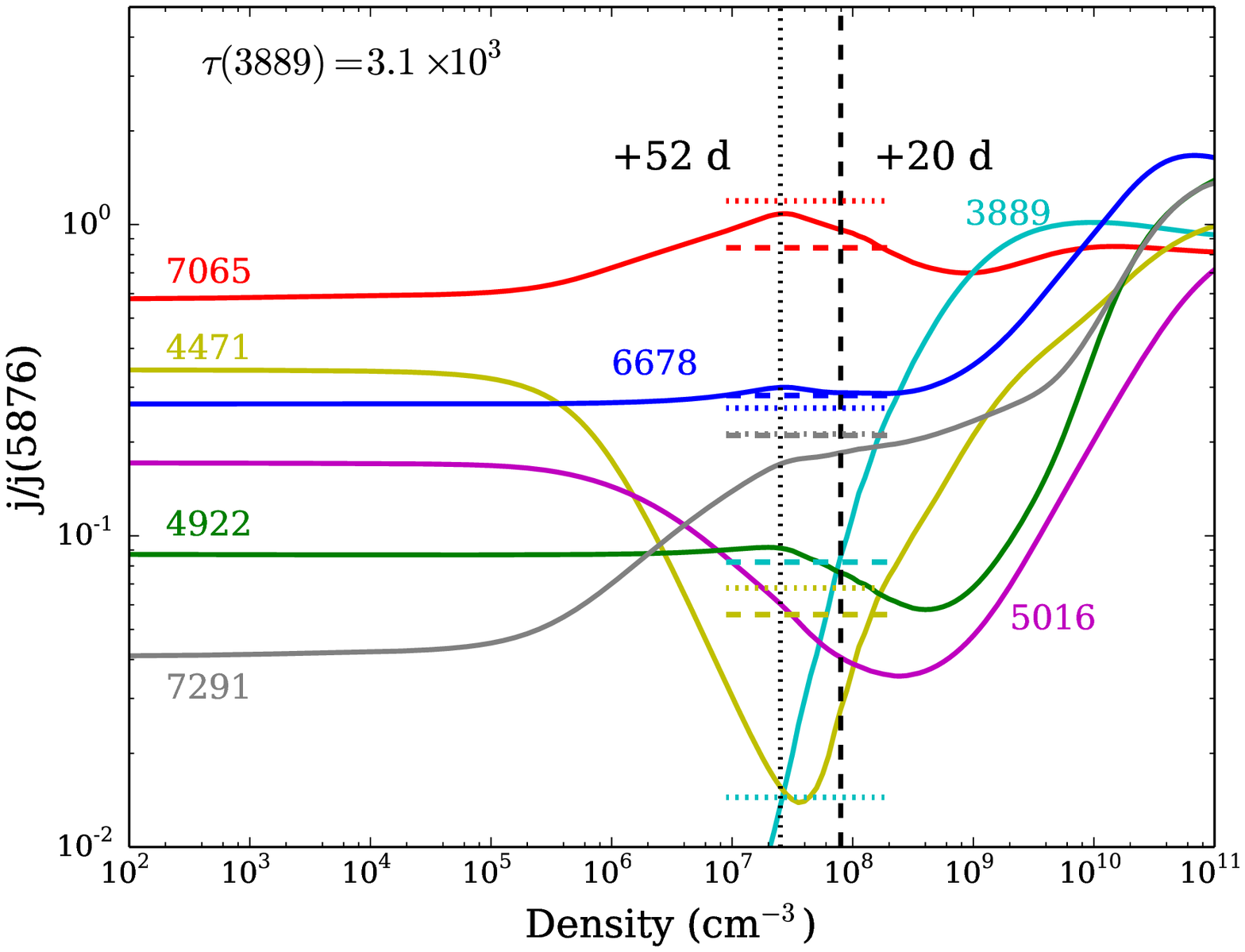}
\includegraphics[width=9cm]{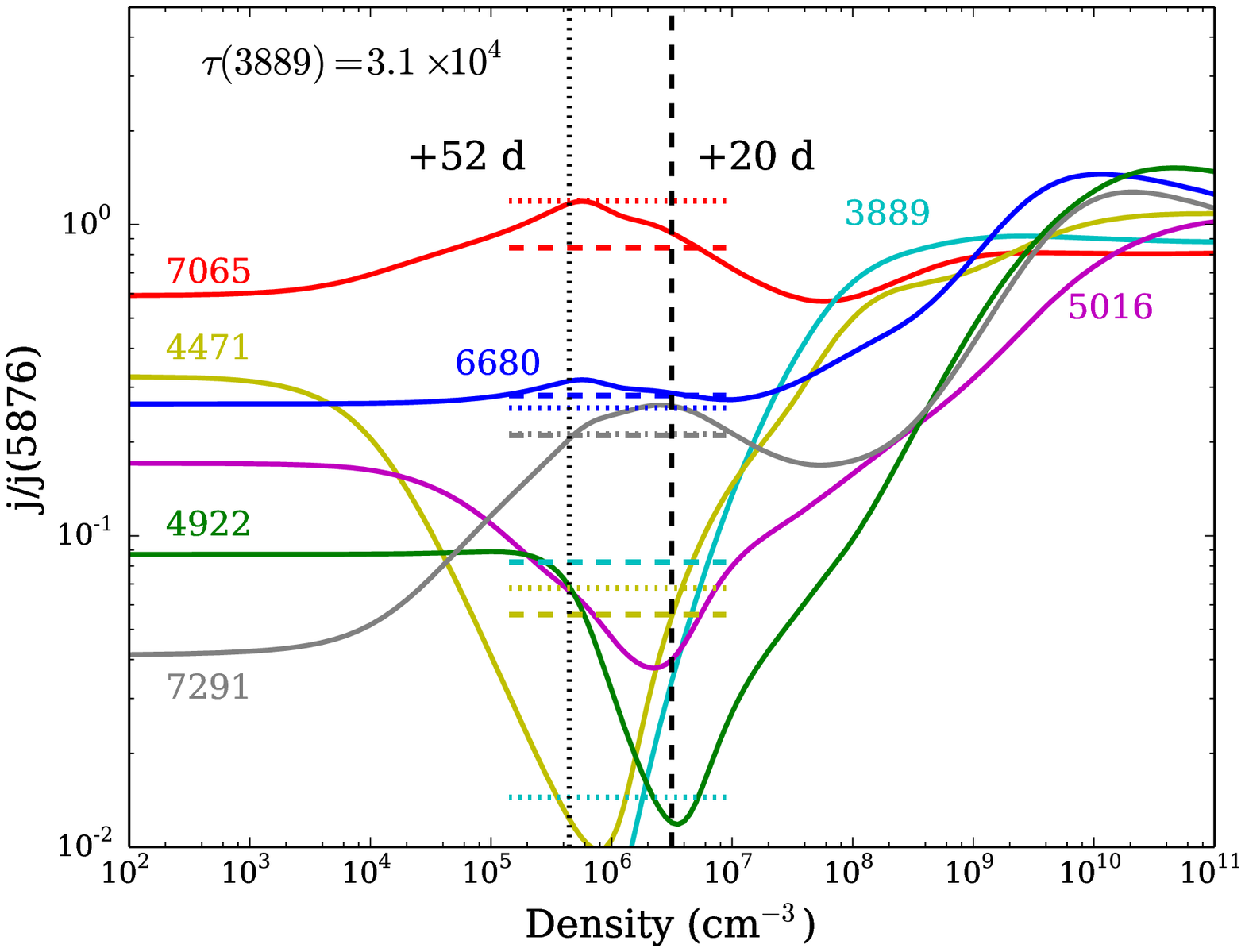}
\caption{Emissivities ($j$) of most important \ion{He}{i} lines relative to the $\lambda~5876$ line. The upper panels show this as function of the optical depth in the $\lambda~3889$ line for two electron densities, $n_{\rm e}=10^6 \text{cm}^{-3}$ and $n_{\rm e}=10^7 \text{cm}^{-3}$, while the bottom panels show the same as a function of the electron density for an optical depth in the \ion{He}{i} $\lambda~3889$ line, $\tau(3889) =3.1 \times 10^3$ (bottom left) and $\tau(3889) =3.1 \times 10^4$ (bottom right). The dashed colored lines give the corresponding observed line ratios for +20 days (dashed lines) and  for +52 days (dotted), while the dashed vertical lines show the density and optical depth with the minimum chi square for these values of parameters. In all cases $T_{\rm e}=7 \times 10^3$ K.}
\label{fig:He_ratios}
\end{figure*}

\begin{figure*}
\centering
\includegraphics[width=9cm]{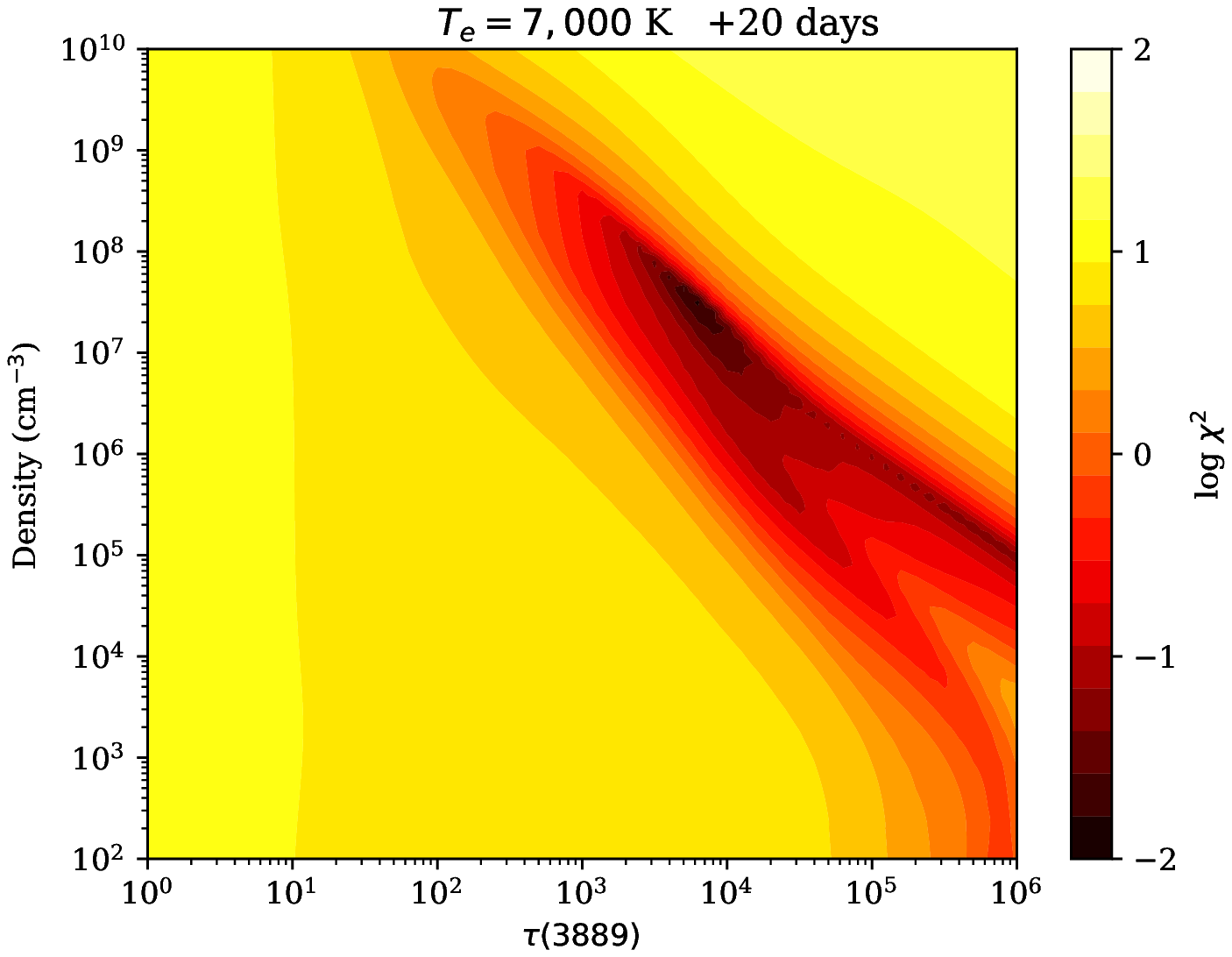}
\includegraphics[width=9cm]{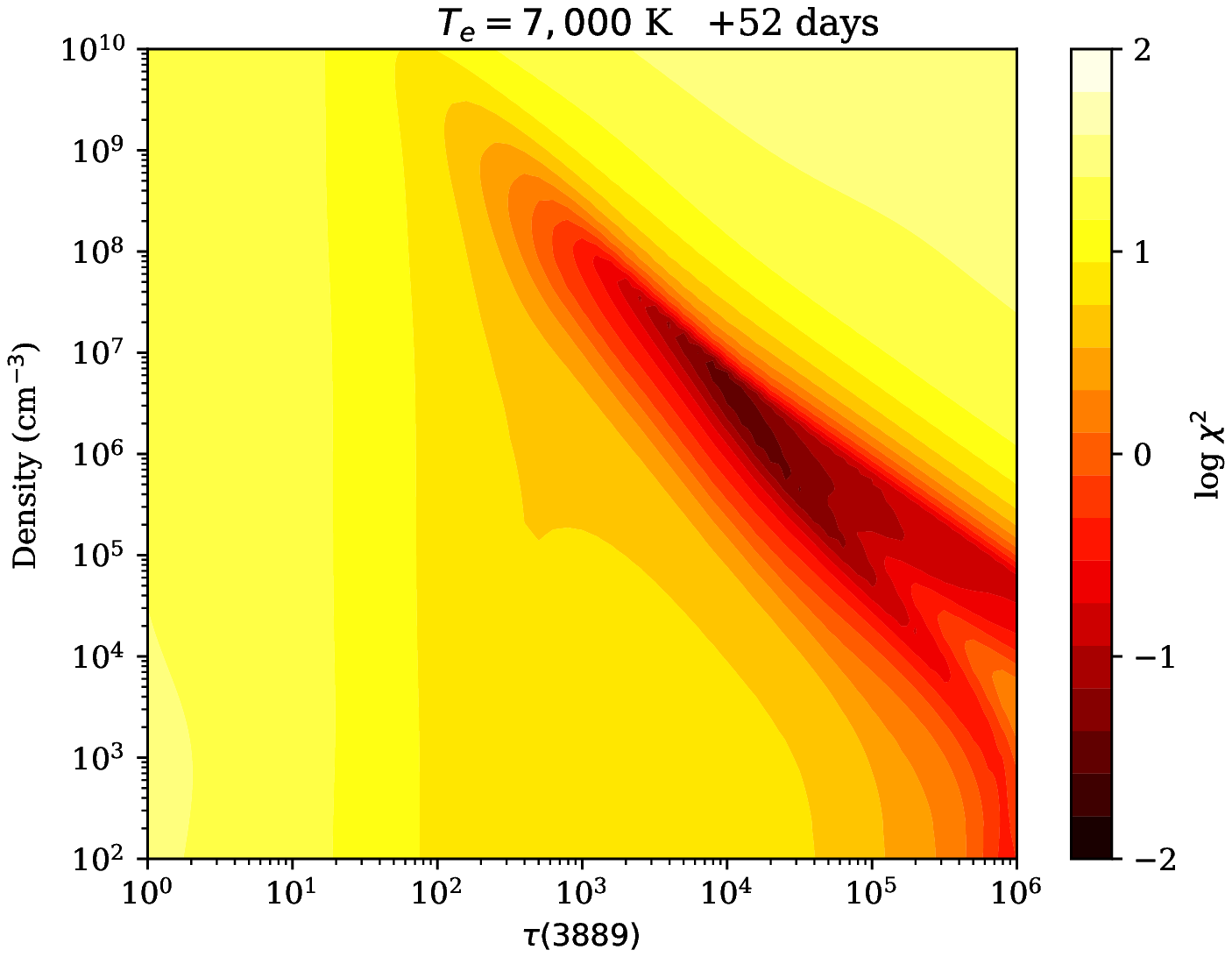}
\caption{Contour plots of $\chi^2$ of the fits at +20 and +52 days as function of the electron density and optical depth in the \protect\ion{He}{i} $\lambda~3889$ line, $\tau(3889)$.  The electron temperature was assumed to be $T_{\rm e}=7 \times 10^3$ K. When calculating $\chi^2$, the \protect\ion{He}{i} $\lambda~7065$, $\lambda~6678$, and $\lambda~7291$, line ratios provide the main constraints, while the $\lambda~3889$ and $\lambda~4471$ lines ratios are used as lower limits since they have strong P-Cygni absorption and weak net emission.}
\label{fig:He_chi2}
\end{figure*}

Table \ref{tab:linesFlux} gives the observed line fluxes of the most important lines relative to the $\lambda~5876$ line at +20 and +52 days. Our main constraints were the $\lambda\lambda~7065/5876$, $\lambda\lambda~6678/5876$, and $\lambda\lambda~7291/5876$ ratios. These were all well determined to $\lesssim 20 \%$. There were also uncertain (but still very useful) estimates from the $\lambda~3889$ and $\lambda~4471$ lines. Although both lines had strong P-Cygni absorption, their total net emission still provides important constraints in conjunction with the other ratios. When calculating $\chi^2$ in Figs. \ref{fig:He_ratios} and \ref{fig:He_chi2}, \ion{He}{i} $\lambda\lambda~6678, 7065, 7291$ line ratios provided the main constraints, while \ion{He}{i} $\lambda\lambda~3889, 4471$ line ratios were used as lower limits since the latter lines were strong in absorption but weak in emission.

\begin{table}
    \centering
    \begin{tabular}{lcc}
    \hline
    \hline
     Line & \multicolumn{2}{c}{Flux ratio relative to the $\lambda~5876$ line} \\
     \AA & Day +20 & Day +52 \\
     \hline
     3889 &0.082   &0.014  \\
     4471 &0.056   &0.068  \\
     4922 &0.15\phantom{0}   &0.34\phantom{0}  \\
     5018 &0.21\phantom{0}   &0.22\phantom{0}  \\
     5876 &1.0\phantom{00}  & 1.0\phantom{00} \\
     6678 &0.28\phantom{0}   &0.26\phantom{0}  \\
     7065 &0.84\phantom{0}   &1.19\phantom{0}  \\
     7291 &0.21\phantom{0}   &0.21\phantom{0}  \\
     \hline
    \end{tabular}
    \caption{Line-flux ratios measured from the spectra used in Sect. \ref{sec:emiss}.}
    \label{tab:linesFlux}
\end{table}

We present our results in Fig. \ref{fig:He_ratios}. In the upper panels of Fig. \ref{fig:He_ratios}, the line ratios relative to the $\lambda~5876$ line as function of optical depth, $\tau(3889)$, are shown for two electron densities, $n_{\rm e}=10^6 \text{cm}^{-3}$ and $n_{\rm e}=10^7 \text{cm}^{-3}$, respectively. In the lower two panels, the line ratios are instead shown as a function of electron density for two different optical depths, $\tau(3889) = 3.1\times 10^3$ and $\tau(3889) = 3.1\times 10^4$, and a temperature of $T_{\rm e}=7 \times 10^3$. The modeled line ratios (colored lines) were compared to the observed ratios at +20 and +52 days, which are shown as horizontal dashed and dotted lines, respectively, with each color corresponding to a specific line ratio. In the rest of the paper, the line ratios omit the denominator ($\lambda~5876$) for conciseness, meaning that $\lambda~7065/5076$ ratio will be $\lambda~7065$ ratio, etc.

As seen in the upper panels, the emissivity of the $\lambda~7065$ ratio increased from a low value at small optical depths to a value close to one for $\tau(3889) \gtrsim 10^3$, where it reached a maximum, before decreasing again. The $\lambda~3889$ and $\lambda~4471$ line ratios became strongly quenched as $\tau(3889)$ increased before subsequently becoming large again as the line ratios approach local thermodynamical equilibrium (LTE) at very high $\tau(3889)$. The same qualitative behavior occurred as function of the electron density as shown in the lower panels. The line ratios are therefore useful as a diagnostic of these parameters.

The black vertical lines in this figure show the best fit values of $\tau(3889)$ or $n_{\rm e}$ to the observations at +20 days (dashed lines) and +52 days (dotted lines). However, the best fit values of $\tau(3889)$ and $n_{\rm e}$ depend both on each other and on the electron temperature. In Fig. \ref{fig:He_chi2} we show the $\chi^2$ value of the best fit as function of these parameters at +20 and +52 days for $T_{\rm e}=7 \times 10^3$~K. There is obviously a strong correlation between the best fit values of $\tau(3889)$ and $n_{\rm e}$. Nevertheless, these figures show that the observations strongly indicated high optical depths for the \ion{He}{i} $\lambda~3889$ line, $\tau(3889) \gtrsim 10^3$, as well as a moderately high electron density, $10^5 \lesssim n_{\rm e} \lesssim 10^8 \text{cm}^{-3}$. While we found that the observed line ratios could be fit in a range of $T_{\rm e}$ between $0.5\text{--}15\times 10^{4}$~K, the acceptable range of density and optical depth was not strongly affected by $T_{\rm e}$ in this range: the best fit CSM parameters were always at high density and optical depth. In addition, $T_{\rm e}\approx7 \times 10^3$ K produced the lowest $\chi^2$ values and was roughly in agreement with the estimated BB temperatures at these epochs.

Although we focused on the optical depth of the $\lambda~3889$ line, optical depths of other lines were also larger than one in the region of the $\tau(3889)$--$n_{\rm e}$ plane favored by the line ratios. The optical depths vary depending on the level at which the absorption takes place (e.g., the $\lambda~4471$ line originates from the $2~{^3P}$ level). For instance, the $\lambda~6678$ line, coming from the singlet level $2~{^1P}$, was the line that became optically thick last of the lines considered, but was still marginally optically thick for the best fit parameters.

Fig. \ref{fig:He_ratios} demonstrates why some \ion{He}{i} lines displayed P-Cygni absorptions while others were dominated by emission. At the high optical depths and densities indicated by our analysis, lines from the lower levels are the ones with the strongest emission. This is a well known result of the branching of emission from higher levels at high optical depth \citep[e.g.,][]{2006agna.book.....O}. For example, the $\lambda~3889$ line into the $\lambda~7065$ plus $4.3 \mu$m and $\lambda~10830$, while the $\lambda~4471$ line splits into $1.7 \mu$m plus $4.3 \mu$m and $\lambda~7065$. Therefore, the fact that 'red' lines are the ones showing strong emission is therefore simply explained by the fact that they cannot branch into other lines. Meanwhile, the 'blue' lines are the ones that trap the photons, which then branch into longer wavelengths. Photons scattered by these therefore emerge at longer wavelengths, resulting in P-Cygni absorptions without the strong emission seen in the redder lines.

\section{Relation to other Type Ibn SNe}
\label{sec:results}

Owing to the increased discovery rate from optical surveys, two comprehensive reviews of Type Ibn SNe have recently been published, P16 and H17, which include 16 and 22 objects respectively, with some overlap. In this section, we compare the results of our analysis to these literature samples of Type Ibn SNe.

\begin{figure*}
\centering
\includegraphics[width=\linewidth]{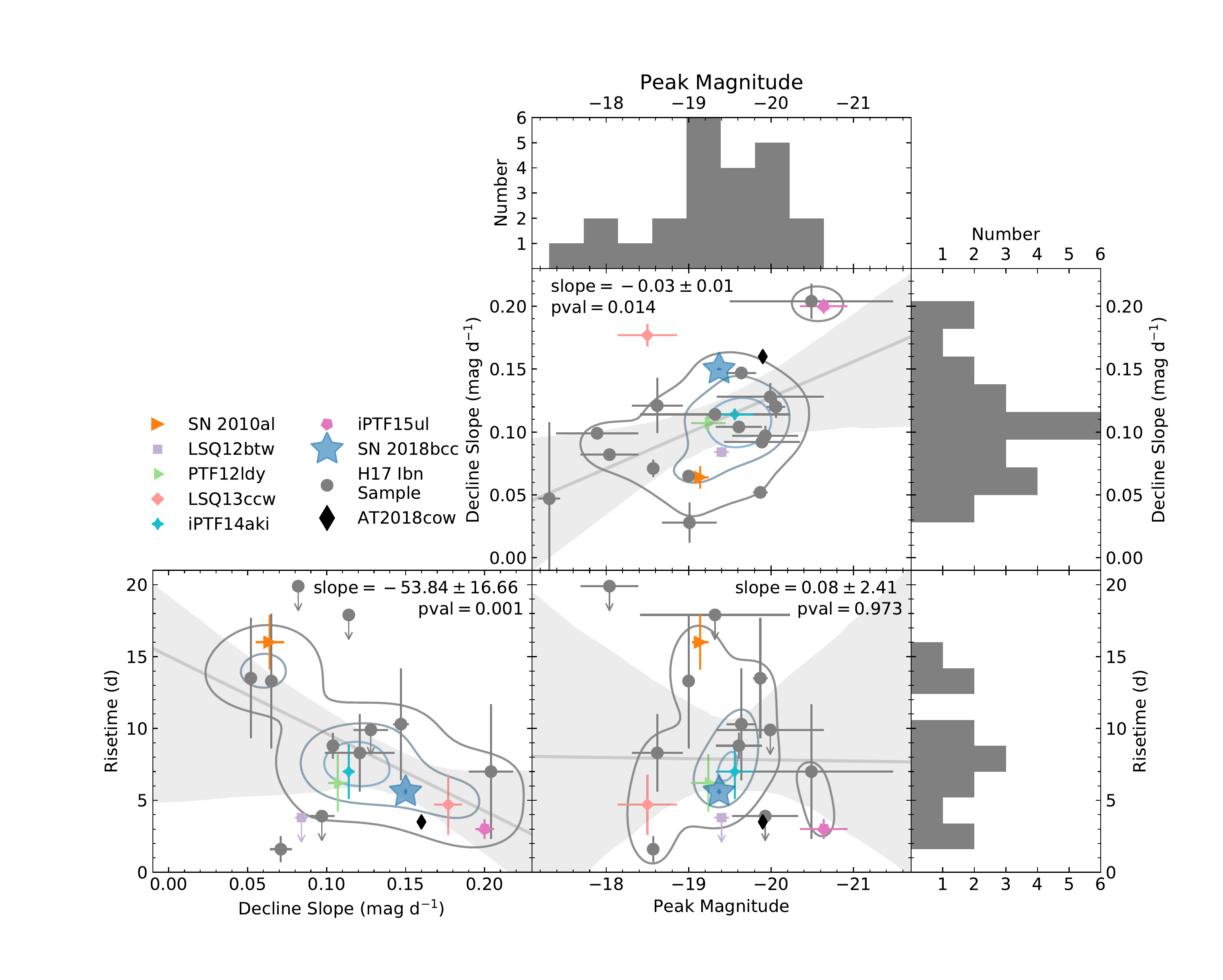}
\vspace*{-10mm}
\caption{Comparison of observational properties of SN~2018bcc with the literature sample of Type Ibn SNe from H17, adapted from their figure 12. SN~2018bcc (blue star) and the comparison sample used in this paper are indicated against the rest (gray circles). The results of an unweighted linear regression are shown in gray lines with the given slope and p-value, with the 90$\%$ confidence intervals indicated by the shaded regions. A density estimate is overplotted for each scatter plot using contours. Based on these results, SN~2018bcc seems to be a typical member of the fast-evolving subtype. The linear regression uncertainties are calculated from the scatter in the data, ignoring the individual errors. The rise time upper limits are not included in the linear regression, density estimates, or the histogram.}
\label{fig:GriffinParComp}
\end{figure*}

Photometrically, SN~2018bcc is a typical fast-evolving example of the Type Ibn class. We show this in Fig.~\ref{fig:GriffinParComp}, which is an augmentation of a figure from H17 (their figure 12), with the rise time, decline slope, and peak magnitude in $r$ band of SN~2018bcc indicated by a blue star. The photometric properties of SN~2018bcc fit within the rest of the sample closer to the faster evolving part of the distribution. Paralleling SN~2010al, the Type Ibn SNe with the best early color information \citep{Pastorello2015a}, SN~2018bcc was blue in color at first, becoming redder, before evolving back to the blue again. A comparable color evolution is also seen for SN~2015U (\addOneNestingLevelEndLink\citealp{Shivvers2016SNSupernova}; H17). Similar to the findings in P16, and as also discussed in several papers on other individual Type Ibn SNe, we find that the radioactive decay of \element[][56]{Ni} alone cannot explain the bolometric properties of SN~2018bcc and that an additional luminosity input, such as from CSM interaction, is needed.

It is important to note that the correlations seen in Fig.~\ref{fig:GriffinParComp} do not include upper limits, which make the rise time correlations unreliable. Luckily, the correlation between decline slope and peak magnitude is not affected by this source of error. From it, we see that brighter Type Ibn SNe tend to decline slightly faster on average, although there is significant scatter that is further impacted by the uncertain contribution of host extinction.

Spectroscopically, the photospheric velocities of the \ion{He}{i} lines, as measured from the minimum of the P-Cygni profile, are similar to those of other Type Ibn SNe discussed by P16, with values around ${\sim1000}$~km~s$^{-1}$ at 20 to 60 days past peak. Additionally, we also saw evidence of a bump blue-ward of ${\sim}5500$~\AA{} that is likely from a forest of Fe II lines. We also detected emission lines from \ion{Ca}{ii} and \ion{Mg}{ii}. A relatively strong \ion{Mg}{ii} emission feature is often detected in Type Ibn SNe (P16), with varying degrees of relative strength of the \ion{Mg}{ii} lines and potentially the nearby \ion{O}{i} line (Fig. \ref{fig:linesID2}). Our deep late-time spectra suggested an evolution in the \ion{Ca}{ii} and \ion{Mg}{ii} features, with the former becoming much stronger and the latter almost disappearing\footnote{While this evolution was not reported in the samples of P16 and H17, the reason could be the paucity of their late-time spectra.}. This could indicate that the mixed \ion{Mg}{ii} and \ion{O}{i} feature quickly wanes after peak for fast-evolving Type Ibn SNe. Interestingly, LSQ12btw and LSQ13ccw, both very rapidly evolving Type Ibn SNe, show a relatively weak mixed feature of Mg and O located around ${\sim7700}$~\AA{}.

We did not see a transformation of the relatively narrow \ion{He}{i} emission lines into broader lines ($V\gtrsim4000$~km~s$^{-1}$) with strong P-Cygni, as is seen in a few Type Ibn SNe (SN~2010al, ASASSN-15ed, SN~2005la, and SN~2015G). These have sometimes been referred to as transitional Type Ib/Ibn SNe, since their spectra evolve to become more similar to ordinary Type Ib SNe in some cases. Instead, we observed the line profiles of the emission dominated \ion{He}{i} lines become more boxy by day 52, without an obvious strong P-Cygni absorption.

A slight declining trend in the FWHM of the \ion{He}{i} lines is observed for SNe~2006jc, 2014av, and 2002ao, which we did not unambiguously detect in SN~2018bcc. This trend has been interpreted by P16 as a decline in the velocity of the shocked gas regions, perhaps resulting from a steeper density profile of the CSM. We note that, alternatively, a declining FWHM could simply mean that the electron scattering depth decreases and therefore the lines become narrower.

We did not detect strong hydrogen features in our spectra before $+52$~d, although resolved H$\alpha$ was, in fact, detected in the final spectrum with a line profile that was narrower than the \ion{He}{i} lines. H$\alpha$ and other hydrogen features are sometimes seen in Type Ibn SNe and can be similar in strength to the \ion{He}{i} lines.
A relatively strong \ion{Mg}{ii} emission feature is often detected in Type Ibn SNe (P16), with varying degrees of relative strength of the \ion{Mg}{ii} lines and potentially the nearby \ion{O}{i} line (Fig. \ref{fig:linesID2}). Interestingly, the fast-evolving Type Ibn SNe LSQ13ccw and LSQ12btw show relatively weak \ion{Mg}{ii} and \ion{O}{i}. The mixed feature of Mg and O located around ${\sim7700}$~\AA{} evolved in our spectra and became weaker at late times, which could indicate that this feature generally wanes quickly past peak for fast evolving Type Ibn SNe. Finally, we also detected the \ion{Ca}{ii} NIR triplet which has been seen for most Type Ibn SNe that have spectral coverage at these wavelengths a few weeks after explosion.

\section{Discussion \label{sec:discuss}}

So far, we have highlighted the unique observables of SN~2018bcc, modeled its LC and \ion{He}{i} lines, and compared the characteristics of this SN with the expanding literature sample of Type Ibn SNe. In this section, using our results, we propose an explanation for the formation of P-Cygni lines in Type Ibn SNe, which may account for the observational division seen in the latest sample of Type Ibn SNe (H17). Afterwards, we also discuss the powering mechanism and CSM properties of SN~2018bcc. We find that SN~2018bcc, and by extension fast Type Ibn SNe, are particularly fitting candidates for pulsational-pair instability (PPI) SNe. Finally, we summarize the evidence constraining the progenitor of SN~2018bcc and conclude that it may not even be a terminal CC SN explosion.

\subsection{Are there two types of Type Ibn SNe? \label{sec:twotypes}}

As discussed in the introduction, H17 point to the possible existence of two observational subtypes based on the early \ion{He}{i} line profiles. One class shows narrow P-Cygni profiles while the other has He-lines purely in emission\footnote{H17 note that this division may also represent a continuum of Type Ibn SNe with short and long-lived P-Cygni lines.}. In SN~2018bcc, all \ion{He}{i} features showed narrow P-Cygni with velocity minima around ${\sim1000}$~km~s$^{-1}$ at least up to ${+14}$~days. However, in the later spectra, the blue \ion{He}{i} features ($<5000$~\AA) kept their narrow P-Cygni profiles while the redder features such as \ion{He}{i} $\lambda\lambda$~6678,~7065 lost their P-Cygni profiles, which were initially weaker to begin with. Thus, SN~2018bcc cannot cleanly fit into the simple division proposed by H17 based on the presence or lack of the narrow P-Cygni line profiles. This suggests that with higher SNR spectra (especially bluewards of $<5000$~\AA) the picture might be more complicated. The same spectrum can both show and not show P-Cygni profiles, which has also been noted in H line profiles of Type IIn SNe \citep[e.g.,][]{Taddia2013a}. Similar to SN~2018bcc, the spectra of the nearby Type Ibn SN~2015G are also found not to fit this simple division \citep{Shivvers2017}.

H17 suggest that the difference between these two observational classes of Type Ibn SNe can be explained by a difference in optical depth; the ones dominated by emission should be optically thin while the ones showing P-Cygni profiles should be optically thick. The mix of scattering dominated P-Cygni lines and emission dominated lines in the spectra of SN 2018bcc shows that this picture is too simplified. As our modeling in Sect. \ref{sec:HeIFlux} showed, the mix of P-Cygni and emission-dominated \ion{He}{i} lines in SN~2018bcc are a result of the high optical depths and densities in the CSM. In particular, the lines dominated by emission (and lacking a clear P-Cygni at late times), such as the $\lambda\lambda~5876, 6678, 7065$ lines, are all still optically thick. They are emission dominated because they lack other lines to branch into, not because they are optically thin.

So, how are the P-Cygni lines formed? In SN~2018bcc, the line producing region responsible for the P-Cygni profiles is likely to be in the regions of the CSM where $\tau_{\rm e} < 1$ (Sect. \ref{sec:emiss}). The ionization of He is expected to be caused by UV and X-rays produced at the shock deep in the interaction region. However, the ionized region responsible for the electron scattering, as well as the He ionization and recombination, is likely to have a moderate $\tau_{\rm e}$. This is the case for a radiative shock where the optical depth of this region is regulated by a balance between the ionizing flux from the shock and the total recombination rate.  As discussed in \cite{Fransson2014}, $\tau_{\rm e} \propto V_{\rm shock}^3$, but is independent of the mass-loss rate of the progenitor. A $V_{\rm shock} \lesssim 10^4$~km~s$^{-1}$ results in $\tau_{\rm e} \lesssim 10$, which may only correspond to a fraction of the total column density of the CSM. While in this picture most of the emission and electron scattering may be produced by the entire ionized region outside of the shock, the P-Cygni absorption should originate from the outer part of this region where $\tau_{\rm e} < 1$.

X-rays penetrating increasingly further out into the P-Cygni-producing region will lead to stronger emission features, which may fill in the absorption. This can help explain the gradual shift to more emission-dominated line profiles we see in the spectral sequence of SN~2018bcc and of Type Ibn SNe in general.

There is some observational evidence for the existence of X-rays from deep in the interaction region in Type Ibn SNe: both SN~2006jc and SN~2010al had X-ray detections. The X-ray LC of SN~2006jc was seen to peak ${\sim}100$d after the optical \citep{Immler2008}. SN 2010al had only a marginally significant X-ray detection with Swift/XRT \citep{Ofek2012a}, but there were indications that the peak occurred well after maximum optical light in this case also.

A late X-ray peak is also expected on theoretical grounds. Assuming that He is singly ionized and a solar abundance of metals by mass (which dominate the X-ray absorption), the ratio of photoelectric absorption at 1 keV to electron scattering is $\sim 10^3$, decreasing approximately as $E^{-3}$. Then, the fact that $\tau_{\rm e} \approx 5-10$ means that the X-ray optical depth can be vary large below 10 keV. This estimate only includes the radial optical depth through the ionized region and not the neutral medium outside this zone or scattering within the ionized zone. The actual X-ray absorption may therefore be even larger. As the column density ahead of the shock decreases with time, the optical depth should also decrease and X-rays may leak out, predominantly at high energy. The clumpiness of the medium can speed this up further (by making lower column-density lines of sight possible). Thus, X-rays that are initially trapped would leak out at later times, leading to an X-ray LC that peaks after the optical, such as the one that was observed for SN~2006jc.

In an alternative scenario, H17 suggest that a viewing angle effect in nonspherically symmetric CSM, such as a torus, can be responsible for the two spectral classes. In this case, edge-on and face-on views, respectively, show or hide the P-Cygni profile. Although we propose a different explanation based on theoretical and observational grounds, viewing angle effects in asymmetric CSM could still play a role.

\subsection{Rise time and powering mechanism of SN~2018bcc}
To date, SN~2018bcc is the best example of a fast Type Ibn SN with a well-sampled early LC. As Table~\ref{tab:risetimes} shows, rise times and the shapes of the rising LCs of fast Type Ibn SNe have been uncertain. We find that both the rising LC of the previous best candidate, iPTF15ul, and of SN~2018bcc are compatible with a t$^2$ power law (as measured by our formulation). We emphasize that the form of the fit is empirical, not physically based, and was chosen for the simple fact that it is able to reproduce the LC shape. Nevertheless, the fact that both rapidly evolving Type Ibn SNe with good early coverage, iPTF15ul and SN~2018bcc, are well fit by a similar exponent is interesting, and calls for further study of whether other fast Type Ibn SNe also have similar early LC shapes. In interaction-powered SNe with optically thick CSM, the rising LC is likely created by shock breakout occurring inside the CSM \citep{Chevalier2011}. The shape of this rise powered by diffusion is then sensitive to the density profile and geometry of the CSM \citep[see e.g.,][]{Ofek2010}. Therefore, regularity in the rise of some Type Ibn SNe could be an indication of regularity in these conditions.

Similar to many other Type Ibn SNe, the rapid evolution and brightness of SN~2018bcc cannot be reproduced by a pure \element[][56]{Ni}-powering scenario, without luminosity input from another powering mechanism. In the case of Type Ibn SNe, which are defined by the spectroscopic signs of interaction with a He-rich CSM, a natural explanation for this additional luminosity is CSM interaction. Using a model of shock-interaction powering from \citet{Chatz2012} we showed that a reasonable mass of dense CSM around the SN could provide the required additional luminosity. We caution that our simplified modeling was performed via visual inspection of the fit for physically plausible values of the model parameters, since we were mainly interested in whether or not CSM interaction could be the potential powering scenario. More sophisticated modeling on the entire Type Ibn SN sample is required to learn more about the actual values of the model parameters and its limitations.

\subsection{CSM interaction properties of SN~2018bcc}

In this section, we discuss whether a consistent picture of the CSM around SN~2018bcc has emerged from our study. In Sect. \ref{sec:modeling}, we modeled the CSM properties of SN~2018bcc using both LCs and spectra. One way to compare the results from LCs and spectra is to estimate the mass-loss rate implied by the analysis or model. However, our LC modeling was based on the work of \citet{Chatz2012}, and is designed for slowly evolving high-mass SNe, such as SLSNe. While the simplifying assumptions in this model may be appropriate for SLSNe, the same assumptions are less appropriate for rapidly evolving low-mass SNe, such as Type Ibn SNe. One particular issue is that the diffusion time of the model will be overestimated for low-mass SNe, causing the model to prefer fits with artificially lower CSM and ejecta masses \citep[see][]{Clark2020}. Thus, we only use the model to indicate that CSM interaction is a plausible powering mechanism and not to derive the detailed properties of the CSM. For the latter, we rely on our spectral modeling instead. Nevertheless, we estimate mass-loss rates from LC and spectral-based methods to act as a comparison.

\citet{Moriya2016} use the peak bolometric luminosities and rise times of Type Ibn SNe to estimate their explosion properties and CSM densities. For the ejecta density profile assumed in their work, the CSM density parameter D given by their equation 4 requires the observables $L_p$, peak luminosity, and $t_d$, the rise time. Using the values from their work and inserting the peak luminosity and rise time of SN~2018bcc, we obtained $D\approx 3.4 \times 10^{15}$~g~cm$^{-1}$, in perfect agreement with the other Type Ibn SNe in their sample. With the wind velocity ($v_w$), we can also estimate the mass-loss rate of the progenitor as  $\dot M \approx 4\pi v_w D$, obtaining

\begin{equation}
    \dot M \approx 0.07 \left(\frac{v_w}{1000~\text{km}~\text{s}^{-1}}\right) \Msun~\text{yr}^{-1}.
\end{equation}

Similarly, our CSM interaction LC models can also be used to estimate the mass-loss rate. Assuming a wind-like CSM, the mass-loss rates can be derived using $\dot M \approx 4\pi\rho_{csm} R_p^2 v_w$, where $R_p$ is the radius of the constant photosphere assumed in the model and $\rho_{csm}$ is the CSM density in g~cm$^{-3}$. For CSM model 2 from Sect. \ref{sec:modeling} (Fig. \ref{fig:arnett}), we obtain

\begin{equation}
    \dot M \approx 0.06 \left(\frac{v_w}{1000~\text{km}~\text{s}^{-1}}\right) \Msun~\text{yr}^{-1}.
\end{equation}

Using a typical wind velocity suitable for a WR star ($v_w \sim 1.5 \times 10^3$~km~s$^{-1}$), we obtained ${\sim}0.1~\Msun$~yr$^{-1}$ for both estimates. Although for the second estimate, the previously noted bias for a lower CSM mass of our model for the case of SN~2018bcc implies that the density, and hence the mass-loss rate, could be higher.

Looking at the spectra, our modeling of the line profile at +52 days in Sect. \ref{sec:HeIFlux} resulted in an optical depth $\tau_{\rm e}=8.6 (T_{\rm e}/ 10^4 \ {\rm K})$. In principle this provides an estimate of the mass-loss rate. Assuming a wind profile for the outflow with velocity $V_{\rm exp}$

\begin{equation}
 \tau_{\rm e} = \int_{R_{\rm in}}^{R_{\rm out}} \frac{\dot M \kappa_{\rm e}}{4 \pi V_{\rm exp} r^2 }dr = \frac{\dot M \kappa_{\rm e}}{4 \pi V_{\rm exp}} \left(\frac{1}{R_{\rm in}}-\frac{1}{R_{\rm out}}\right) .\
\end{equation}

For a gas dominated by singly ionized He $ \kappa_{\rm e} \approx 0.1$. The outflow velocity is uncertain. We therefore scaled it to $V_{\rm exp} = 1000 \kms$, which is in the observed range $600 - 2000 \kms$  for the H$\alpha$ blue shift and the terminal velocity of the P-Cygni profile (Sect. \ref{sec:HeIFlux}). We have only weak constraints on the inner and outer radii of the ionized region, $R_{\rm in}$ and  $R_{\rm out}$. If we assume that $R_{\rm in} \ll R_{\rm out}$, only $R_{\rm in}$ is important. As a rough estimate we assumed that $R_{\rm in} \approx V_{\text{exp}} t \approx  10^3 \kms \times 46 \ {\rm days} \approx 4.0 \times 10^{14}$ cm. Putting all this together we arrived at

\begin{equation}
\dot M \approx 0.68 \left(\frac{V_{\rm exp}}{1000 \kms}\right)^2 \left(\frac{T_{\rm e}}{10^4 \ \rm K}\right)^{-1/2}  \left(1-\frac{R_{\rm in}}{R_{\rm out}}\right)^{-1} \ \Msun~\text{yr}^{-1}. \
\label{eq:mdotspec}
\end{equation}

When considering this estimate, we note that there are large uncertainties in several of the parameters. While the opacity and outflow speed are probably within a factor two, the inner radius and thickness of the region in particular are highly uncertain. On one hand, a thinner shell would increase the mass-loss rate, while a lower expansion velocity would lower the mass loss considerably. Therefore, Eq.~\ref{eq:mdotspec} should mainly be considered as an order of magnitude estimate.

The \ion{He}{i} optical depths and densities inferred from our modeling in Sect. \ref{sec:HeIFlux} can also provide an estimate of the mass-loss rate. However, because the observed \ion{He}{i} lines all come from highly excited levels, estimates of the optical depth require estimates of the fractional populations of these levels. These in turn require detailed modeling of the \ion{He}{i} spectrum, including photoionization, recombination, and collisional processes, i.e. a complete modeling of the SN spectrum, which is beyond the scope of this paper. Even without this, we find that both spectra and LC-based estimates of the mass-loss rate are very high, in the region associated with eruptive mass-loss in massive stars \citep{Smith2014c}.

Our spectral modeling offers further insight into the properties of the CSM. Modeling of the \ion{He}{i} lines in SN~2018bcc has shown that optical depth and electron density are degenerate, and the best fits to the observed line fluxes are all located in a region of high density and optical depth (Sect. \ref{sec:emiss}). A high optical depth is also supported by our modeling in Sect.~\ref{sec:HeIFlux}. As we described in Sect. \ref{sec:twotypes}, optically thick CSM can help explain the evolution of P-Cygni lines in SN~2018bcc, and probably in Type Ibn SNe.

Based on these results, we conclude that SN~2018bcc has a dense and optically thick CSM environment, with a progenitor that experienced a very high mass-loss rate. Such a high mass-loss rate, combined with the suggestion that the CSM around Type Ibn SNe is in a spatially confined shell \citep{Moriya2016,Hosseinzadeh2017}, suggests a different mechanism than ordinary steady-state line-driven winds as the likely creator of the CSM around Type Ibn SNe. A potential candidate is eruptive mass-loss via some mechanism such as wave-driven super-Eddington winds \citep{Quataert2016} or the PPI \citep{Woosley2007,Woosley2017}, which are both expected to lead to the formation of CSM shells. While there are few observational predictions from the acoustic mechanism, \citet{Fuller2018} predict a mass-loss rate ${\sim}0.1 \Msun~\text{yr}^{-1}$, in line with our estimates for SN~2018bcc. We discuss the PPI in detail in Sect. \ref{sec:PPI}.

Narrow ($\lesssim 1000$~km~s$^{-1}$) but clearly resolved H$\alpha$ was seen in the final spectrum of SN~2018bcc, although it was blueshifted by $300 \kms$ (Appendix, Fig. \ref{fig:mgca}), representing the systematic redshift uncertainty of SN~2018bcc. It did not show a boxy profile unlike the \ion{He}{i} lines in this spectrum. In order to explain the appearance of H$\alpha$, we consider two cases. In the first case, the ejecta or CSM are catching up with slower and more H-rich material ejected previously. However, in this case one would expect similar widths for the H and He lines, contrary to the observations. Alternatively, this H-rich material may be located much further out, and is instead being photoionized by X-rays produced in the interaction region. This would explain the lower velocity of the H-rich material. We discuss this possibility further in the next section. We note that a successful model of CSM interaction in SN~2018bcc would need to both explain the peculiarities of the \ion{He}{i} lines profiles, as well as the late-time rise of narrower H$\alpha$.

\subsection{Comparison with pulsational pair-instability supernova models \label{sec:PPI}}

\citet[][hereafter W17]{Woosley2017} discuss the possibility that Type Ibn SNe may be the result of PPI SNe. These have an origin in high-mass stars with He-core masses between $30 \text{--} 64$ $\Msun$, corresponding to ZAMS masses of 70--140 $\Msun$ (with a large uncertainty due to unknown mass-loss rates). As such massive stars approach the pair instability region, they undergo violent nuclear flashes that send strong enough pulses to eject any H-envelope, as well as parts of the He core. The number of pulses and the duration between each pulse varies strongly with the He-core mass. The durations range from hours to a few thousand years, with more energetic pulses leading to a larger interval (W17). Interaction between shells ejected in these pulses, or if energetic enough an ejection in itself, can create SN-like transients referred to as PPI SNe.

Both the luminosities and the shape of the LCs of the resulting SNe vary strongly with the He-core mass. In particular, the peak luminosity and the rising part of the LC vary strongly between different models, which is a result of both the energy in the pulses and the time interval between pulses. Unlike pair instability SNe, PPISNe leave behind a massive ($>30 \Msun$) remnant that (eventually) undergoes CC, either directly forming a black hole, or possibly exploding as an SN. Due to the altered composition of the remnant, a neutrino-driven explosion is thought to be unlikely (W17). Hence, the possible contribution of a final CC SN is ignored in the proceeding discussion.

\begin{figure}
\centering
\includegraphics[width=1.0\linewidth, angle=0]{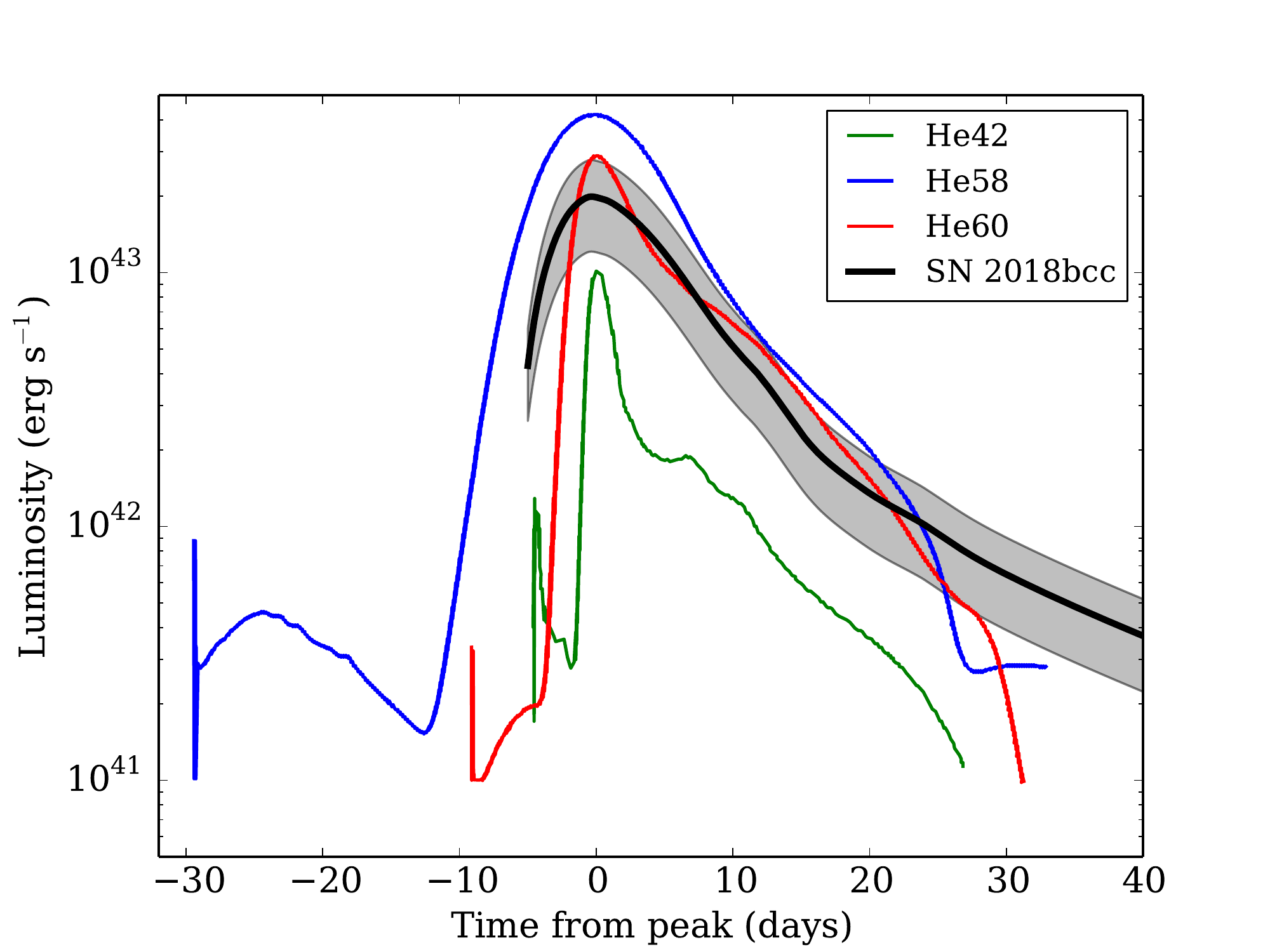}
\caption{Comparison of bolometric LC of SN~2018bcc, including errors (gray, shaded region), with the bolometric LCs of the He42 (green), He58 (blue) and He60 (red) models from W17.}
\label{fig:WoosleyPPISNComp}
\end{figure}

Although suggested before, there have been few direct comparisons of model PPISN LCs with Type Ibn observations. We have therefore used our bolometric LC for a comparison with the W17 models of He stars. The models are named so that the He-core mass of a model is indicated by its number; for example, He50 is a 50~$\Msun$ He-core mass model. In Fig. \ref{fig:WoosleyPPISNComp} we compare our bolometric LC with the LCs of the He42, He58, and He60 models from W17 (his fig. 6). It is important to stress again that the LCs of the models vary significantly with He-core mass in a nonlinear way. In fact, the LCs of the high-mass models (He58 and He60) are formed in a different way than the LC of the low-mass model (He42).

The He42 model is included as an example at the lower mass-range for a PPISN and was among the models discussed by W17 as a possible candidate for a Type Ibn SN. The duration of pulsations in the lower mass-range models is only a few days, and therefore the energetics are lower. In total, He42 ejects $2.65 \Msun$ of material in several pulses that power the LC seen in Fig. \ref{fig:WoosleyPPISNComp}. The luminosity of the He42 model is a factor of $\sim 5$ lower than that of SN 2018bcc. The total radiated energy is similarly lower; $\sim 3.9 \times 10^{48}$ ergs for the He42 model while it was $\sim 2.1 \times 10^{49}$ ergs for SN~2018bcc. The effective temperature at the peak was $23 \times 10^4$ K in the model compared to the observed $\sim 15 \times 10^4$ K. These differences with SN~2018bcc also apply to the He40 and He44 models.

For intermediate mass models He48--He54, the LCs all have multiple peaks as a result of colliding shells emitted within months to a few years before collapse. Hence, despite the fact that the peak luminosities of these models are in the range of that of SN~2018bcc, the overall widths of these LCs are far too broad. Therefore, these models are excluded for SN~2018bcc.

Higher mass models, such as He58 and He60, have fewer but more energetic pulses and a much longer duration between the first and second pulse, on the order of thousands of years. However, similar to the lower mass models, the later pulses occur on a time scale of days and the ejected shells merge on a short time scale resulting in a smooth LC. Therefore, as can be seen in Fig. \ref{fig:WoosleyPPISNComp}, the LCs of the He58 and He60 models are in better agreement with the observed bolometric luminosity of SN~2018bcc and are consistent with it given the limitations of the models. These LCs do not show the first pulse, which occurred thousands of years prior and is predicted to be optically faint.

The He58 and He60 models have total radiated energies of $\sim 4.4 \times 10^{49}$ ergs and $\sim 1.9 \times 10^{49}$ ergs, respectively. This is more similar to the total radiated energy for SN~2018bcc, $\sim 2.1 \times 10^{49}$ ergs, than the lower mass He42 model. In the latter pulses, they expel $4.8\Msun$ and $2.0\Msun$ of material for the He58 and He60 models, respectively\footnote{The first pulse expels ${\sim}10 \Msun$, which should now be located at several hundred to a few thousand AU.}. They also have effective temperatures in the range of $8,000\text{--}12,000$ K, which is consistent with the  observed temperatures for SN~2018bcc (Fig.~\ref{fig:BBtemps}).

The ejecta velocity is another important diagnostic. For the low-mass models, the photospheric velocities are expected to be in the range of $2000\text{--}4000 \kms$.  Meanwhile for He58 and He60, the resulting velocity only varies between $2500-3000 \kms$. The He lines in SN~2018bcc have FWHM velocities in the range of the PPISN models (Sections \ref{sec:Helines} and \ref{sec:HeIFlux}). Although, the observed bulk velocity may be lower due to CSM interaction (Sect. \ref{sec:HeIFlux}).

While this comparison is very interesting, there are a number of caveats when comparing observations to the models. The largest uncertainty in the observed LCs comes from the bolometric correction estimate, due to the lack of UV and NIR data. We have tried to correct for this when calculating the bolometric correction, but there are likely to be considerable uncertainties, as shown by the shaded area in Fig. \ref{fig:WoosleyPPISNComp}. Nevertheless, the observational uncertainties are relatively minor compared to those of the models.

Comparing different model calculations \citep[e.g., W17;][]{Yoshida2016,Marchant2018,Leung2019}, there is broad agreement in terms of energies, number of pulses, and ejected masses. However, there are considerable quantitative differences between models in the literature for the same He-core mass. Furthermore, small differences in He-core mass can cause large difference in the LCs for the same model calculations, since LC behavior is very sensitive to the He-core mass. These one-dimensional models cannot account for instabilities caused by shell interactions, which are likely to be important and to modify the hydrodynamic structure. When comparing with observations, additional uncertainty comes from the fact that the LCs of the W17 models are only bolometric, with no color information, and the only opacity source included when calculating the LCs is electron scattering and line absorption is neglected. Since all of these factors are important, differences by a factor of two or more are not unexpected.

Nevertheless, there are many agreements between the observational properties of SN~2018bcc and theoretical predictions for PPISNe. In addition to what has been mentioned previously, the appearance of the narrow H$\alpha$ line at late times can be readily accounted for; it may originate from a more distant shell produced by an earlier ejection, possibly the first pulsation, which ejected most of the hydrogen envelope. We inferred a relatively low metallicity for the host galaxy of SN~2018bcc, in agreement with observations by \citet{Hosseinzadeh2019} for several other Type Ibn SNe. Line-driven stellar winds are less efficient at low metallicity, and hence massive stars can more easily survive past the main sequence with larger envelope masses, which are necessary for PPISNe. The late decline of SN~2018bcc and several other fast Type Ibn SNe which seem to lack a radioactive decay tail is also readily explained in the PPISN scenario\footnote{Although, the ultimate fate of the remnant in producing a CC SN is unknown.}. Finally, the most important test for the PPISN scenario is that the ejected material should not show enhanced abundances from elements resulting from advanced nucleosynthesis. Despite the fact that a few Type Ibn SNe evolve to show more ejecta-like spectra, we did not detect evidence of SN ejecta or enhanced abundances of elements from advanced nucleosynthesis in the spectra of SN~2018bcc, matching previous observations of a few other fast Type Ibn SNe \citep[see e.g.,][]{Pastorello2015}. Thus, some luminous and fast Type Ibn SNe with these characteristics are attractive candidates to be PPISN explosions, even if other Type Ibn SNe may not be. Additionally, the eventual CC of the remnant, which we did not consider in this discussion due to the large unknowns, may be able to produce ejecta-like spectra seen in some Type Ibn SNe.

One of the predictions of PPI is the depletion of oxygen. Thus, a lack of strong oxygen in late-time spectra (like what we saw in SN~2018bcc), can also be a litmus test for PPI in Type Ibn SNe. Interestingly, the late-time spectra of SN~2006jc also lacked oxygen, similar to most other Type Ibn SNe. Even in SN~2015G, which is one of the few Type Ibn SNe with more ejecta-like late time spectra, the oxygen line flux is much lower compared to ordinary SE~SNe \citep{Shivvers2017}.

Observations that can further test the PPISN scenario for Type Ibn SNe include a study of early and late-time brightening in Type Ibn SNe, as either pulsations or shell interaction can lead to pre- and post-SN brightening. However, since the time between pulsations and the duration of pulsations varies widely from hours and days to thousands of years, a statistical study is necessary. On the modeling front, LCs from two-dimensional models with realistic stellar envelopes and evolution are needed for a fruitful comparison to observations.

\subsection{Progenitor of SN~2018bcc}
Type Ibn SNe have been suggested to be the CC of massive Wolf-Rayet (WR) stars exploding into dense and He-rich CSM. Like for other Type Ibn SNe in literature, the CSM around the progenitor of SN~2018bcc would have to have been ejected via a mechanism with a very high mass-loss rate. The SN also seems to be powered by CSM interaction, with little \element[][56]{Ni}. In this progenitor scenario, a normal CC SN explosion could be hidden beneath the CSM-interaction-powered LC. Even if it comes out that there is less \element[][56]{Ni} than expected, one possible way to explain a low \element[][56]{Ni} mass combined with a CC SN could be via fallback onto the central compact object. However, looking at the available observational evidence, we cannot conclude that this progenitor scenario is the only possibility.

SN~2018bcc was located in a low-metallicity environment. It seems to be powered by CSM interaction, with little or no \element[][56]{Ni}, and was a rapidly evolving and fading SN with a relatively low total radiated energy of $\approx 2.1 \times 10^{49}$~erg (low for an interaction powered SN). The spectra neither showed direct evidence of nucleosynthesis nor fast, SN-like, ejecta (velocities ${\sim} \text{10,000}~\kms$). Therefore, it is also worth exploring the possibility that SN~2018bcc, and perhaps other fast Type Ibn SNe, are not terminal CC SNe. As discussed in Sect. \ref{sec:PPI}, if the progenitor of SN~2018bcc was a massive star in a low-metallicity environment, the LC and interaction properties (from the spectra) could be the result of the collision of CSM shells ejected via some mechanism, such as the PPI. In this alternative scenario, the progenitor of SN~2018bcc is a very massive star whose remnant may even collapse to form a blackhole, either directly or as a failed SN \citep[e.g.,][]{Gerke2015}.

In the case of PS1-12sk, the only Type Ibn SN to have exploded outside of an elliptical galaxy with deep limits on the star formation rate, \citet{Sanders2013} and \citet{Hosseinzadeh2019} evoked the possibility that the progenitor was a white-dwarf system. For SN~2018bcc, the rapidly declining LC and lack of \element[][56]{Ni} also makes thermonuclear explosions an unlikely progenitor scenario. As noted in the study of Type Ibn SN host environments by \citet{Hosseinzadeh2019}, all Type Ibn SNe may not even share a common progenitor. Various progenitor systems can create interaction powered SNe with strong \ion{He}{} emission lines. It seems that in the case of SN~2018bcc, either the CC of a WR star or nonterminal ejection into a dense, He-rich CSM, are possible progenitor scenarios. PPI is one plausible mechanism that can give rise to either. Ultimately, the nature of the progenitors of fast Type Ibn SNe remains unknown.

\subsection{AT2018cow}
As seen in Fig. \ref{fig:GriffinParComp}, SN~2018bcc is a fast Type Ibn SN that is otherwise ordinary. Remarkably, the photometric properties of the enigmatic transient AT2018cow are not that different from SN~2018bcc or other fast Type Ibn SNe. Furthermore, AT2018cow developed post-peak \ion{He}{i} emission features that look similar to many other Type Ibn SNe \citep{Perley2019}, along with several other spectral similarities. These similarities were highlighted in a recent paper by \citet{Fox2019}, who discussed the CSM interaction scenario for AT2018cow in detail (however see \citealp{Wang2019}). We note that in our basic modeling, SN~2018bcc could be mostly \element[][56]{Ni} powered for characteristic ejecta velocities $\gtrsim 0.1c$. Although we lack any evidence for such relativistic ejecta in SN~2018bcc, and thus discount this possibility, AT2018cow had an early broad feature with just such a high velocity \citep{Perley2019}. If CSM interaction can explain the observables of AT2018cow, then, just like other Type Ibn SNe, it might be a viable PPISN candidate. In this scenario, very high velocity ejecta seen at early times could be coming from the CC of the remnant shortly after the cessation of pulsations, while the CSM interaction in the ejected shells can explain other observables. If AT2018cow is indeed similar to fast Type Ibn SNe, further study of this nearby enigmatic object might also illumine the mystery surrounding Type Ibn SNe, or vice versa.

\section{Conclusion} \label{sec:concl}
We studied the rapidly evolving Type Ibn SN~2018bcc and have obtained the best constraints on the combined rise time and shape of the rising LC for this type of fast Type Ibn SN. It has a rise time of $5.6^{+0.2}_{-0.1}$~days in the restframe with a rising power law index close to two. Photometrically and spectroscopically, SN~2018bcc seems to be a typical but fast-evolving member of the Type Ibn SN class.

Modeling of the bolometric LC suggests that \element[][56]{Ni} cannot be the sole powering-mechanism for this SN, even when accounting for gamma-ray leakage. An additional luminosity input is needed, which we showed with our simple modeling could readily be provided by CSM interaction. Our spectral modeling suggests that the \ion{He}{i} line profiles had strong electron scattering wings and thus are likely not representing the expansion velocity. Furthermore, the observed \ion{He}{i} line flux ratios suggest optically thick CSM and a high electron density, $10^5 \lesssim n_{\rm e} \lesssim 10^8 \text{cm}^{-3}$. In addition, the \ion{He}{i} lines were all optically thick even at $+52$ days.

The CSM around SN2018bcc is likely produced by a high mass-loss rate of ${\sim}0.1~\Msun$~yr$^{-1}$, which suggests an alternative formation mechanism, such as eruptive mass loss. We also did not find strong evidence of CC SN signatures. As a result, SN~2018bcc can be the result of a CC SN interacting with a He-rich CSM ejected in an eruptive episode, or possibly the interaction of CSM shells ejected via an eruptive mechanism, such as the PPI mechanism. We show that, given all of the available evidence, SN~2018bcc, and, by extension, other fast Type Ibn SNe, are viable PPISN candidates.

We have found that the simple spectral classification scheme proposed by H17, based on the evolution of the narrow P-Cygni profiles of \ion{He}{i}, does not neatly fit SN~2018bcc. We saw \ion{He}{i} features both with and without narrow P-Cygni profiles in the same spectra, with P-Cygni profiles present in the blue part and emission dominated features in the red part of the spectrum. Furthermore, their explanation that optical depth effects can explain the discrepancy is found to be too simplified, since all lines are still optically thick. Instead, the branching of emission from higher levels in a He atom at high optical depths is likely responsible for the creation of P-Cygni profiles. Lines that lack branching possibilities become emission dominated, for example, redder lines like $\lambda~7065$.

The nature of Type Ibn SNe remains an ongoing mystery. There are even possible associations to extreme events such as AT2018cow, which has been shown to share many observational similarities to the class. We reiterate that more sophisticated modeling on a large sample of Type Ibn SNe is required to answer questions regarding their progenitors, CSM conditions, and powering mechanisms. As of today, observations are lacking to undertake this type of population study for Type Ibn SNe (often done for other SN subtypes). Primarily, there is a lack of early-time photometric observations that can constrain the shape of the rise and peak of these objects, and virtually no color information for the early LCs except for a few specific cases (see e.g., H17). The lack of early time data also makes estimating the explosion epoch difficult, which is a necessary step for LC modeling. Since Type Ibn SN LCs decline rapidly, we also often lack nebular spectroscopy that might be able to tell us more about the ejecta and SN properties, such as expansion velocities. Finally, there has been no systematic exploration of the CSM interaction in Type Ibn SNe in radio and X-rays, a topic that has been deeply studied for Type IIn SNe, for instance.

This paper was written with the aim of reporting early results from the ZTF survey with an example of a unique rapidly evolving SN discovered shortly after the start of the public ZTF survey in March 2018. As the case of SN~2018bcc shows, robotic surveys like the ZTF can be used to discover a sample of well-observed candidates required to undertake a sample study of Type Ibn SNe.

\begin{acknowledgements}
We gratefully acknowledge support from the Knut and Alice Wallenberg Foundation and the Swedish Research Council. Based on observations obtained with the Samuel Oschin Telescope 48-inch and the 60-inch Telescope at the Palomar Observatory as part of the Zwicky Transient Facility project. Major funding has been provided by the U.S National Science Foundation under Grant No. AST-1440341 and by the ZTF partner institutions: the California Institute of Technology, the Oskar Klein Centre, the Weizmann Institute of Science, the University of Maryland, the University of Washington, Deutsches Elektronen-Synchrotron, the University of Wisconsin-Milwaukee, and the TANGO Program of the University System of Taiwan. The Oskar Klein Centre is funded by the Swedish Research Council. This research is partially based on observations made with the Nordic Optical Telescope, operated by NOTSA at IAC using ALFOSC, which is provided by the IAA. Some of the data presented herein were obtained at the W.M. Keck Observatory, which is operated as a scientific partnership among the California Institute of Technology, the University of California, and NASA; the observatory was made possible by the generous financial support of the W.M. Keck Foundation. Some of the data presented herein were obtained with the Liverpool Telescope operated on the island of La Palma by Liverpool John Moores University in the Spanish Observatorio del Roque de los Muchachos of the Instituto de Astrofisica de Canarias with financial support from the UK Science and Technology Facilities Council. This work is partly based on observations made with DOLoRes@TNG. This research made use of Astropy, a community-developed core Python package for Astronomy (Astropy Collaboration, 2018) IRAF, is distributed by the National Optical Astronomy Observatories, which are operated by the Association of Universities for Research in Astronomy, Inc., under cooperative agreement with the National Science Foundation. The python version which was used, PyRAF, is a product of the Space Telescope Science Institute, which is operated by AURA for NASA. SED Machine is based upon work supported by the National Science Foundation under Grant No. 1106171. We would like to thank E. Ofek for his comments on the manuscript.
\end{acknowledgements}

\bibliographystyle{aa}
\bibliography{ZTFIbn_ads}

\begin{appendix}
\section{Constructing the bolometric LC}
\label{app:bololc}
There are three standard methods of constructing a bolometric LC for CC SNe. In the simplest method, the SED is integrated directly (typically using trapezoidal integration), as much as observations permit, and the resulting flux is converted to a quasi-bolometric LC that only accounts for the light in the observed part of the SED. This method is the best when an abundance of observations across the electro-magnetic (EM) spectrum exists for an object.

In the second method, the direct integrated quasi-bolometric LC is extrapolated into the IR and UV by using a combination of fits to the SED (commonly using a blackbody) to estimate parts that are outside of the observed range. Typically a correction is then applied to the blackbody fit to account for heavy line blanketing present in the UV, which is usually not sampled by the observations \citep{Lusk2017}. This integrated flux can be converted to a bolometric luminosity using the known distance to the SN.

A third method involves calculating a bolometric correction to the observed SED when observations are limited. These corrections are calculated from another property, usually color, and calibrated using well observed (and nearby) SNe so that a simple parametrized relation between color and the bolometric correction is empirically calculated \citep[see e.g.,][]{Lyman2013}. However, such a parametrized relation does not yet exist for Type Ibn SNe. As a result, creating a pseudo-bolometric LC for a Type Ibn SN often entails obtaining photometry across the EM spectrum.

\subsection{Steps for SN~2018bcc}
In order to create the bolometric LC, we first used our spectra to create a quasi-bolometric LC via direct integration. We used the common range from $4000-8500$~\AA{} for direct integration, but we had to remove the LT spectrum as it does not extend beyond $7500$~\AA. However, since this spectrum is very close in time to the SEDM spectrum (1.5 days in the restframe), we did not lose much information as a result.

We then extended this region to the UV and NIR by using our BB fits. We integrated the BB fits from $3300-4000$~\AA{} to include the UV up to the left edge of the U-band and from $8500$~\AA{} to effectively infinity to include the NIR. Following other authors \citep[e.g.,][]{Lyman2013} we cut off the SED for wavelengths shorter than 2000~\AA, since there should be strong line blanketing there. We then interpolated from zero flux at 2000~\AA{} to the value of the BB fit at 3300~\AA{} (the left edge of the typical U-band), as most optical LCs of SNe extend to this range, and show it to be well fit by a blackbody \citep[see e.g.,][]{Lyman2013}.

\begin{figure}
\includegraphics[width=\linewidth]{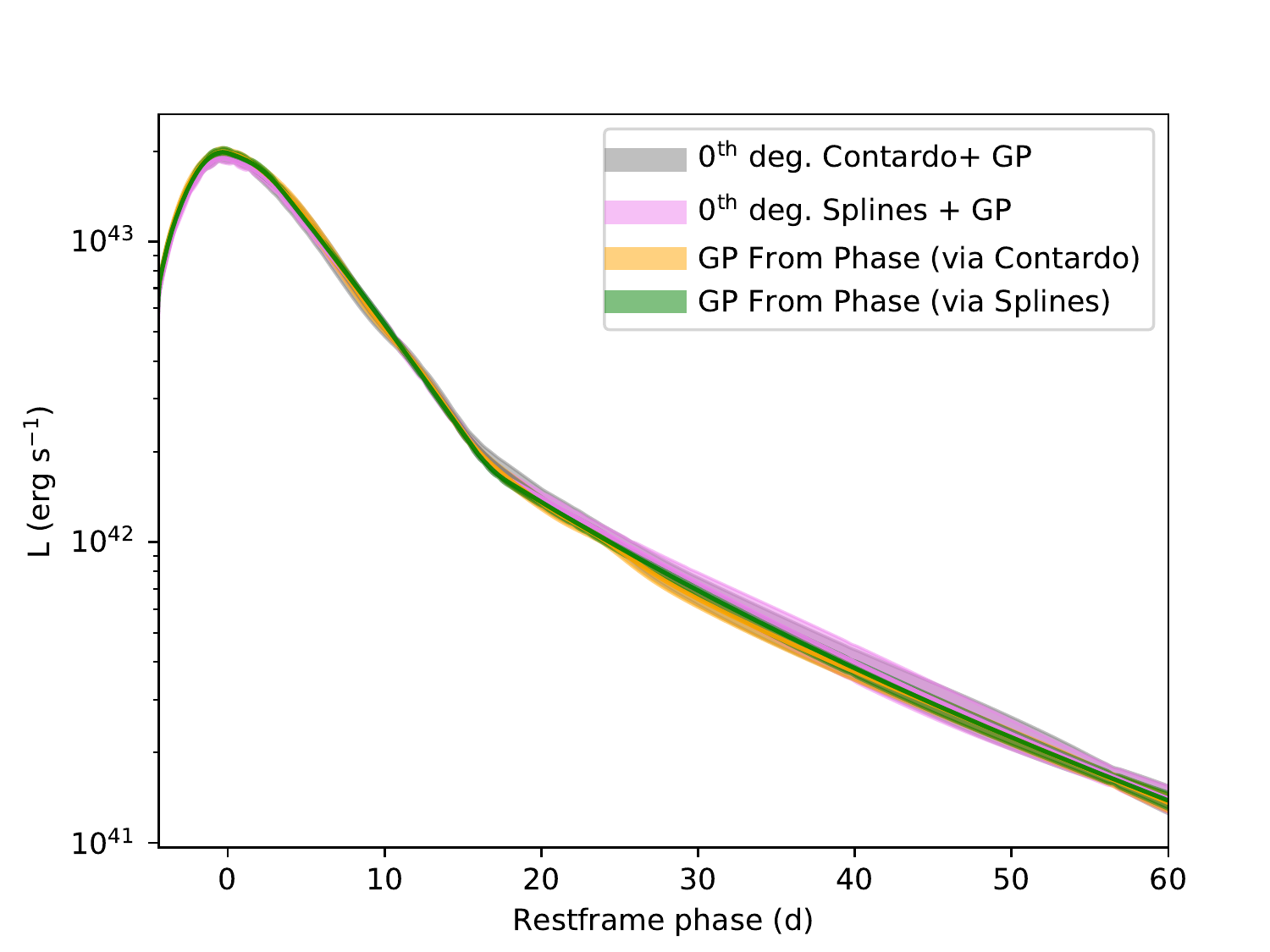}\\
\includegraphics[width=0.96\linewidth]{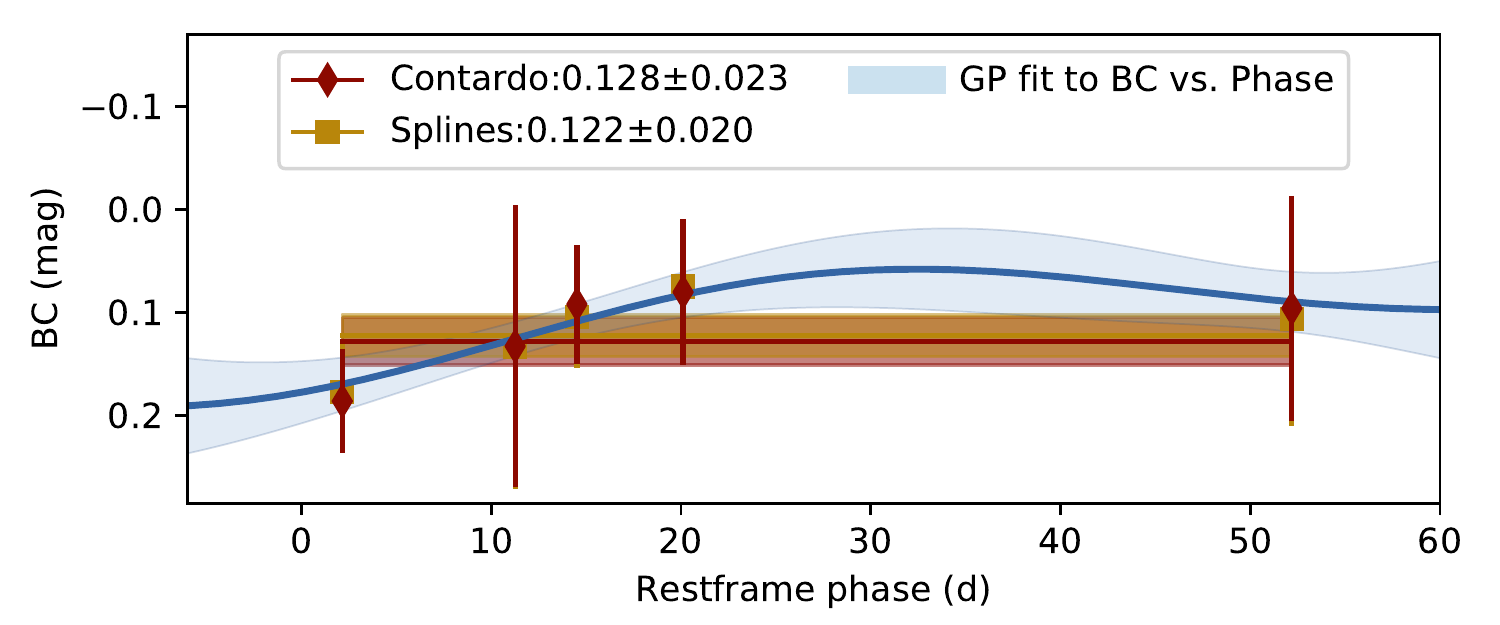}
\caption{\textbf{Top:} Bolometric LC of SN~2018bcc. Results from methods discussed in Sect. \ref{sec:bolo} are plotted with different colors and agree within the uncertainties. \textbf{Bottom:} Fits to the bolometric correction (BC) as a function of phase, obtained via methods discussed in Sect. \ref{sec:bolo} and Appendix \ref{app:bololc}. Resulting bolometric LCs from these methods are plotted in the top panel.}
\label{fig:bolo}
\end{figure}

We converted the integrated flux from this exercise into bolometric magnitudes and calculated a bolometric correction to the $r$-band absolute magnitude LC at the epochs of the spectra. Next, we employed the secondary method and find a simple linear relation between the bolometric corrections we calculated and the $g-r$ color at the same epochs, following the approach of \citet{Lyman2013}. The relation between the $g-r$ band and the bolometric correction is plotted in Fig. \ref{fig:bolo}. We find that a flat (zeroth order) linear fit best describes the relation between the bolometric corrections and the $g-r$ color at the same epochs; a bolometric correction $\approx -0.12 \pm 0.02$~mag. We also tested using first- and second-degree polynomials to characterize this relation and find that our results do not change.

Since a zeroth order correction is preferred using the $g-r$ color, we also looked for a relation between phase and bolometric correction using the interpolated $r-$band LC, which yielded the same result. Additionally, we used GP regression to learn the relationship between phase and bolometric correction and calculated a bolometric LC this way. The resulting bolometric LC is very similar to using a zeroth order bolometric correction as a function of color or phase. Thus, we used our simple bolometric correction to calculate the bolometric LC, as illustrated in Fig. \ref{fig:bolo}.

Finally, we also tested the effects of varying our assumptions when deriving the bolometric correction from the spectra. These tests included using just a simple BB fit, as well as varying the assumptions in the UV, and still found that the resulting bolometric LCs were all similar to our desired accuracy.

\newpage
\onecolumn
\section{Figures}

\begin{figure*}[hb]
\centering
\includegraphics[width=18cm]{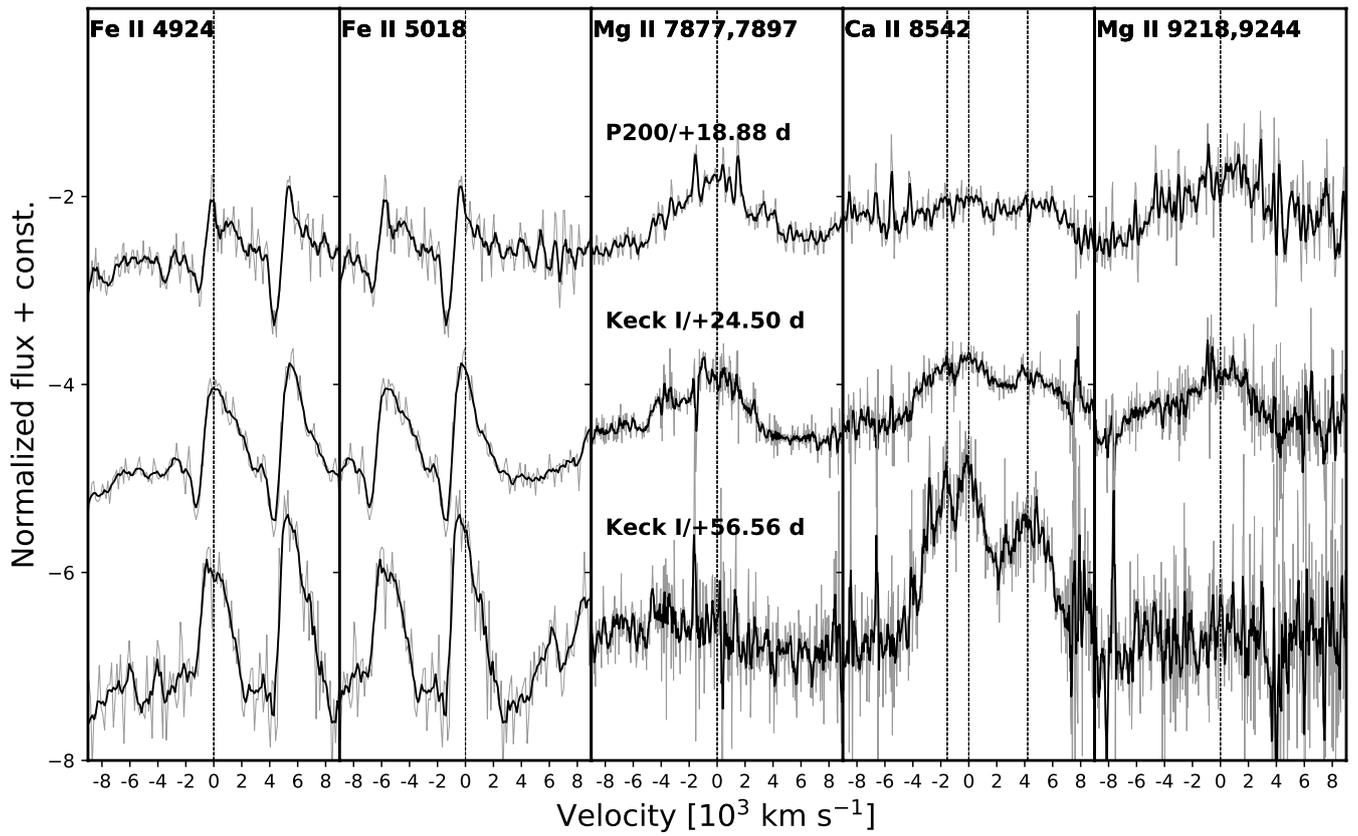}
\caption{Line evolution of prominent \protect\ion{Ca}{ii}, \protect\ion{Mg}{ii} lines and H$\alpha$. While the \protect\ion{Mg}{ii} lines fade by $+56 \text{d}$, the \protect\ion{Ca}{ii} features become stronger with time. H$\alpha$ has been blueshifted by $300$~km~s$^{-1}$ to match the line centroid. \label{fig:mgca}}
\end{figure*}

\section{Tables}
\newpage

\begin{deluxetable}{llllll}
\tabletypesize{\scriptsize}
\tablewidth{0pt}
\tablecaption{Spectral observations \label{tab:spectra}}
\tablehead{
\colhead{JD-2,458,000}&
\colhead{Phase\tablenotemark{a}}&
\colhead{Telescope}&
\colhead{Instrument}&
\colhead{Range}&
\colhead{Resolution\tablenotemark{b}}\\
\colhead{(days)}&
\colhead{(days)}&
\colhead{}&
\colhead{}&
\colhead{(\AA)}&
\colhead{($R$)}}
\startdata
233.83 & 2.19	& P60	& SEDM	 &  3575$-$8629 & {$\sim$}100\\
235.54 & 3.80	& LT	& SPRAT	 &  3782$-$7521 & {$\sim$}350\\
243.50 & 11.28	& NOT	& ALFOSC &  3333$-$9112 & {$\sim$}360\\
246.92 & 14.50	& P200  & DBSP	 &  3198$-$9878 & {$\sim$}1423\\
252.89 & 20.11	& Keck  & LRIS 	 &  2897$-$9679 & 922 $\pm$ 34\\
287.00 & 52.18	& Keck  & LRIS 	 &  2900$-$9674 & 905 $\pm$ 26\\
\enddata
\tablenotetext{a}{Restframe days since peak.}
\tablenotetext{b}{Resolving power $R=\lambda/\Delta\lambda$. Values without error bars are the nominal $R$ at the central wavelength for the instrumental set-up used, obtained from the webpage of the respective instrument. The $R$~values with uncertainties are the "seeing-corrected" versions measured from sky spectra at an average effective wavelength of ${\sim}5900$~\AA.}

\end{deluxetable}

\begin{deluxetable}{llll}
\tabletypesize{\scriptsize}
\tablewidth{0pt}
\tablecaption{\label{tab:linesFit} FWHM and P-Cygni absorption velocities inferred from fitting different lines present at $t>+14\,\rm{d}$. Velocities are spaced with a comma and refer to the fits made on the $+14\,\rm{d}$, $+20\,\rm{d}$ and $+52\,\rm{d}$ spectrum (P200, first and second Keck spectrum), respectively and refer to the atomic line reported in the first column. Systematic redshift uncertainty of ${\sim}300\kms$ applies while typical measurement uncertainties are ${\sim}$200--300~$\kms$ based on noise.}
\tablehead{
\colhead{Line} &  
\colhead{P-Cygni maximum absorption} &
\colhead{Terminal velocity (BVZI)} &
\colhead{FWHM} \\
\colhead{(\AA)} &
\colhead{($\rm{km}\,\rm{s^{-1}}$)} &
\colhead{($\rm{km}\,\rm{s^{-1}}$)} &
\colhead{($\rm{km}\,\rm{s^{-1}}$)}}
\startdata
\ion{He}{i}  3889 &  925,  953, 1350 & 1914, 1830, 1618 & 1830, 1860, 1040$^\text{a}$ \\
\ion{He}{i}  4471$^\text{b}$&  800,  845, 1174 & 1550, 1750, 1850 & 1020, 1380, 1790$^\text{a}$ \\
\ion{He}{i}  4922$^\text{c}$&  834, 1087, - & 1400, 1590, - & 1107$^\text{a}$, 704, 2920$^\text{a}$ \\
\ion{He}{i}  5016$^\text{d}$& 1157, 1221, - & 1745, 1954, - &  755 , 721, 2600$^\text{a}$ \\
\ion{He}{i}  5876 & 1000, 1020, -    & 1600, 1356, -	   & 3000, 3020, 3000 \\
\ion{He}{i}  6678 &  990, 1070, -    & 1470, 1428, -	   & 1950, 2080, 1650 \\
\ion{He}{i}  7065 & 1060, -   , -    & 1645, -   , -	   & 2520, 2100, 1871$^\text{a}$ \\
\enddata

\tablenotetext{a}{Uncertain measurements due to the lower SNR of the spectrum at those wavelengths}
\tablenotetext{b}{Possibly contaminated by \ion{Mg}{ii} $\lambda$~4481 line.}
\tablenotetext{c}{Possibly contaminated by \ion{Fe}{ii} $\lambda$~4924 line.}
\tablenotetext{d}{Possibly contaminated by \ion{He}{i} $\lambda$~5048 and \ion{Fe}{ii} $\lambda$~5018 lines.}

\end{deluxetable}
\end{appendix}
\end{document}